# 75 Years of the Wavefunction:
## Complex-Dynamical Extension of the Original Wave Realism and the Universal Schrödinger Equation

A.P. KIRILYUK*

Solid State Theory Department, Institute of Metal Physics
36 Vernadsky Avenue, 03142 Kiev-142, Ukraine

ABSTRACT. Following Max Planck's hypothesis of quanta (quant-ph/0012069) and the matter wave idea of Louis de Broglie (quant-ph/9911107), Erwin Schrödinger introduced, at the beginning of 1926, the concept of wavefunction and proposed an equation which it should obey, in compliance with the experimentally measured positions of atomic energy levels. Though always endowed with a realistic undular interpretation by its father, the wavefunction of a quantum object could not really be considered as its "matter wave" because of arising fundamental difficulties related to Schrödinger equation linearity and has been provided, until now, with only abstract, formally probabilistic interpretation of scholar quantum mechanics. In this paper we show how and why the resulting "mysteries" of usual theory are consistently solved only within the unreduced, dynamically multivalued description of the underlying, essentially nonlinear interaction process (quant-ph/9902015, quant-ph/9902016), without any artificial, inconsistent modification of the Schrödinger equation itself. The latter is rigorously derived instead as universal expression of unreduced interaction complexity. The causal, totally realistic wavefunction is obtained in this intrinsically complete theory as the dynamically chaotic (causally probabilistic) intermediate state of a system with interaction performing its dynamically discrete transitions ("quantum jumps") between its localised, incompatible "realisations" ("corpuscular" states). Causal wavefunction is a physically real unifying element of any unreduced, complex-dynamic interaction process and therefore can be extended, together with the Schrödinger equation it obeys, to any higher level of the world dynamics and universe in the whole. We outline some important applications of the obtained causally complete description of quantum behaviour, such as genuine quantum chaos, quant-ph/9511034-36, and related realistic quantum devices, physics/0211071, and emphasize the fundamental difference of the proposed dynamically multivalued theory from appearing dynamically single-valued imitations of "causality" and "complexity" in certain external, abstract reformulations of conventional quantum mechanics and unitary substitutes for the unreduced dynamic complexity. The causally extended wavefunction concept, representing the unified essence of unreduced (multivalued) complex dynamics, provides a clear distinctive feature of realistic science, absent in any its unitary imitation.

*Address for correspondence: A.P. Kirilyuk, Post Box 115, 01030 Kiev-30, Ukraine. E-mail: kiril@metfiz.freenet.kiev.ua.

CONTENTS

# 1. Introduction: Unaccomplished wave realism of Erwin Schrödinger and its causal completion in the universal science of complexity

Three founding advances of quantum theory, Max Planck's energy quanta hypothesis [1], "matter wave" introduction by Louis de Broglie [2], and the final synthesis of quantum dynamics by Erwin Schrödinger (emphasised in the present paper), have a number of impressive similarities in the apparently accidental ways of their emergence, further development and views of their authors. A common feature of their appearance is striking unexpectedness and surprising speed at which those truly "breakthrough", revolutionary ideas have come to their authors and were immediately put by them into an appropriate form of scientific result.[*] The dramatic birth of the quanta hypothesis and Planck's constant during a two-months period in October-December 1900 [1,3] is "replayed" 23 years later [2,4] when, according to de Broglie, "a big light suddenly appeared in my mind" giving rise to the matter wave idea, which was fixed, within two months, in his three autumn notes of 1923. And two years later the same type of sudden and contradictory revelation of a completely new truth strikes the 38-year-old professor of the University of Zürich, Erwin Schrödinger [5], who specifies, in November 1925 - January 1926, the ideas of de Broglie's "wave mechanics" in a mathematically more complete form of his *Wellenmechanik* (now called "quantum mechanics") including the wavefunction and famous "Schrödinger equation" it obeys. The decisive article containing this main dynamic relation of (nonrelativistic) quantum mechanics was presented in its final form to *Annalen der Physik* on the 27th of January 1926 [6], after dramatic Schrödinger's remake of the article in search for the right problem solution [5], and then was quickly followed (in February-June 1926) by its other three parts [7-9] and a couple of additional articles, including derivation [10] of previously proposed rival formalism of "matrix", or "quantum", mechanics of Heisenberg-Born-Jordan (1925) from Schrödinger's "wave mechanics". During this very short period of just a few months a major part of standard fundamental and applied "quantum mechanics" was thus created by a mature scientist who had, until then, only marginal, occasional interest [11] in a generally popular and quickly developing field of quantum phenomena (here again one finds a resemblance with Planck's and de Broglie's discoveries).

The well-known difficulties with physical interpretation of the wavefunction appeared practically together with the basic Schrödinger formalism and remain unsolved, within any scholar approach, until today, which creates an ever growing feeling of frustration (see e. g. [12]), especially on the background of proclaimed practical omnipotence of the same science paradigm. The father of the wavefunction always believed himself in a physically "real wave" it should describe and tried to resolve the fundamental difficulties of this realistic interpretation, in accord with his general holistic *Weltbild* [13] and in a "strangely" persisting opposition [14,15] to the mainstream combination of "quantum" mystification and "mathematical" interpretations. We find here again a close similarity between the "stubbornly realistic" attitudes to quantum world understanding of E. Schrödinger, M. Planck [1] and L. de Broglie [2] persisting through years of their long scientific activity, despite the apparent absence of consistent solution and dominating opposition of the "international scientific community" of abstract-minded scribes. Needless to say, the totally consistent, causally complete understanding of the very basis of the physical world is indispensable *de facto* for any further development of fundamental and practical knowledge, and today yet much more so than at the time of incorruptible quest for truth of Erwin Schrödinger and other realistically thinking founders of the new physics.

---

[*]The high speed and easiness of publication of those *strikingly* novel results of the "new physics" almost immediately after their presentation and despite the existing quite vigorous opposition of approaches and people is a separately important feature of that period and way of scientific creation that was still dominated by "old-fashioned" intrinsic realists who used to be *honest* with respect to both nature and scientific colleagues. Modern domination of fruitless "mathematical physics" inevitably brings quite opposite practice, determined by ambiguous abstractions, crudely biased "circles of friends", traffic of influences, and strict, purely subjective elimination of any "opposite" opinions from any printed source, occurring today even without any scientific explanation. Already because of that kind of "moral climate" in modern scholar science, nothing comparable to the *conceptually* new physics advance of the beginning of the 20th century can emerge today by way of "officially permitted" (and therefore fruitless) scientific publications, whereas such qualitative advance definitely extending the deadly ruptured and abstract knowledge to the "unified diversity" of real world is especially necessary just now because of the critically high and ever growing instrumental, *purely technical* power of science and technology [1].

The origin of difficulties with realistic wavefunction interpretation is related to the basic *linearity* of Schrödinger formalism, as opposed to the *irreducible nonlinearity* of the original wave mechanics of Louis de Broglie (the "double solution" scheme [16-18]) which, however, could not be unambiguously specified, either physically or mathematically. Whereas de Broglie insisted upon essentially nonlinear dynamics behind its externally dominating linear "envelope" [2,4], Erwin Schrödinger preferred to look for a simpler, basically linear origin of the wavefunction properties [14,15], while always retaining the fundamentally *realistic* character of the proposed interpretation. This particular feature of Schrödinger's approach can be considered as manifestation of a general tendency towards internal contradictions in his exceptionally diverse talent and complicated personality [5,13] including that striking combination of intrinsic realism and deceptive "transparency" of conceptually "easy", linear and perturbative, solutions within the canonical paradigm of "mathematical" physics (in this respect, his scientific approach is close to that of Max Planck [1,3]).

The *causally complete* extension of canonical quantum mechanics, intrinsically unified with the causally extended versions of "relativity" (special and general), "field theory" and "particle physics", was recently developed as a result of the *unreduced*, universally nonperturbative analysis of underlying *interaction* processes [19-22] (see also [1,2]). The unreduced interaction analysis naturally leads to the universal, reality-based concept of *dynamic complexity* of *any* real interaction process (including causally specified, universal notions of "nonlinearity", "chaoticity", "self-organisation", etc.) that just provides, due to its *conceptual* novelty, a totally causal solution of all canonical "quantum mysteries", unified around the consistent, *realistic* interpretation of the Schrödinger wavefunction, which does not involve its speculative "interpretation" or any artificially imposed, *technical* ("nonlinear" or "stochastic") addition to the conventional formalism. This truly *first-principles* approach deals exclusively with the "main", driving *dynamics* of a configurationally simple, *a priori* structureless system of two *physically real*, interacting entities ("protofields") and shows how *all* the observed entities (particles, waves, interactions, atoms, etc.) and their properties progressively *emerge* in the natural, unreduced development of this interaction process, without any artificial, tricky addition of "inexplicable" and abstract "postulates", "laws of nature", "fundamental principles", etc. Instead, all the known postulates, laws, and principles of the canonical science are *consistently derived*, in their *causally extended* version, from the single starting "postulate" of the theory, which simply specifies the physical, qualitative nature of interaction components (protofields) for our universe. Known particular laws are but (usually reduced) manifestations of the *single*, causally justified, *universal order*, specified as *conservation, or symmetry, of complexity* (the latter being defined according to the results of interaction analysis) [19]. Such wholeness and realism of this new approach, tentatively called "universal science of complexity" and "quantum field mechanics" (in its applications to lowest, quantum levels of the world dynamics), would certainly be considered as the desired "final goal" by such "incorrigible" realists and unifiers as Planck, de Broglie, and Schrödinger. Note that the obtained crucial advance with respect to stagnating canonical abstractions and mysticism is due to the explicitly derived, *conceptual* novelty, the *dynamic redundance (multivaluedness) and entanglement* phenomenon, resulting from the unreduced, universally nonperturbative interaction analysis and giving the consistently specified version of dynamic complexity, with all its universally extended manifestations (chaos, "self-organisation", "adaptability", etc.).

In this article we concentrate on the causally extended wavefunction concept, including its realistic, complex-dynamical interpretation, consistent derivation of the Schrödinger equation (causal "quantization"), its complex-dynamical solutions for an arbitrary real interaction (leading, in particular, to the true quantum chaos [23,24] and consistent quantum measurement description [25]), and universal generalisation of both wavefunction and Schrödinger equation to higher (arbitrary) levels of complex world dynamics [19], expressing the mentioned unified conservation of complexity.

We especially emphasize the *unifying* character and role of the causally specified wavefunction in complex interaction dynamics, as if reflecting the unified diversity of its inventor's talent [5,13], even though he never mentions anything close to our unreduced complexity with respect to quantum mechanics. Indeed, the unreduced interaction analysis shows that the wavefunction is none other than a single, though dynamically random and permanently changing, "transitional state" between different incompatible *realisations* of a real system with interaction and therefore can be considered as the physically real "quintessence", or "conductor", of the whole "orchestra" of multivalued dynamics, both in quantum mechanics and its generalisation to higher levels of complex world dynamics [1,2,19-22]. We show how this special, *irreducibly* complex-dynamical (multivalued) role of the wavefunction explains both its mysteriously dualistic, "unreal" status in all the "officially permitted" (= dynamically single-valued) "interpretations" of quantum behaviour and the absence of this notion in conventional imitations of complexity always restricted to the same, perturbative paradigm of canonical science.

In conclusion, we justify the objective necessity and urgency of a qualitatively big, revolutionary change towards the *unreduced* description of reality in the fundamental physics and beyond (proposed by the universal science of complexity [19]), in relation to actually observed "scientific revolution" effects (T. Kuhn) showing a subjective aspect of knowledge development and critically amplified in this transitional epoch.

## 2. Unreduced interaction dynamics and causal wavefunction emergence

### 2.1. Fundamental dynamic redundance and nonlinear quantum beat process

A major deficiency of the whole conventional science approach, especially clearly seen in the standard quantum mechanics scheme and all its unitary modifications, "interpretations" and development towards "field theory", is the basically "non-interactional", "given" nature of the main concepts and description obtained, in the best cases, by *postulation* of *mathematically* generalised observation results, *without* any *consistent derivation* that should reproduce the *genuine* process of *natural emergence* of the corresponding *real* entities. It is clear that such emergence can only occur as a result of some *interaction* process development considered in its unreduced, non-perturbative version. By contrast, any explicitly perturbative, or "model", "exact-solution", interaction description of the canonical science *cannot* reproduce its *main* result, realistic *structure formation*, and is forced therefore to artificially *insert*, in one way or another, a crudely simplified, abstract "model" of this structure (thus postulating also its main *properties*) and then develop only its minor details or equivalent aspects. In the case of conventional Schrödinger formalism, one uses a combination of known (but always *taken for granted*) postulates of *classical* mechanics and then *adds* some new, *explicitly flawed* "quantum" postulates obtained as direct, unexplained generalisation of experimental observations. It is *not* really surprising (or "unreasonable") that the obtained equation provides solutions that agree, *in combination* with the "inexplicable" postulates, with the experimental results: such agreement was explicitly implanted in the description from the beginning. As for universality of the conventional quantum formalism, it is actually rather limited (even in the micro-, let alone macro-, world phenomena) and can only be due to the underlying universality, or "symmetry", of the *real* world dynamics, which is only superficially "guessed" within the purely *abstract* description of experimental facts, while its true origin remains "mysteriously", and today simply *scandalously*, hidden as far as "objective" and "rigorous" approach of the self-important official *science* of "computer age" is involved (cf. "veiled reality" of B. d'Espagnat [26]). As we shall see later, it is precisely those "mysterious" quantum postulates that actually express, in an *artificially perverted* single-valued form, the *essential dynamic complexity*, or "nonlinearity", of the underlying *real interaction* dynamics, which permits the "main" formalism to remain *technically* linear, a possibility that cannot be extended to higher levels of complexity, where its explicitly observable, "nonlinear" manifestations become so abundant that they cannot be hidden any more within a small number of "rigorous" axioms.

In order to obtain a causally complete description of quantum behaviour and avoid its traditional deficiency, one should start with the *unreduced* analysis of underlying *interaction process* and try to *derive* the observed entities and their properties as a result of real interaction development. It can easily be seen [1,2,19-22] that the *simplest* possible configuration of a system with interaction underlying the *unified* world dynamics and giving rise to all the observed entities is provided by two *a priori* homogeneous, physically real protofields uniformly attracting to each other. One of them, the "gravitational" protofield, or medium, is made of a more rigid, "heavy" and "dissipative" material resembling a dense fluid and plays the role of inert "matrix" to which another, "electromagnetic" (e/m) protofield, or medium, is attracted and forms readily the observed structures due to its much more "movable", "light", "compressible" and "elastic" physical nature. System structure (elementary particles, their interactions, etc.) can appear only due to complex, *multivalued* dynamics of the unreduced interaction process (see below), while the protofield names are due to the fact that they eventually give rise to universal e/m and gravitational particle interactions, respectively. We explicitly *deduce* all those details in our theory [1,2,19-22] and fix as a single, unavoidable postulate only the qualitative, physical origin of the protofields specifying the "electro-gravitational" type (quality) of our world. Since every world structure is obtained as developed perturbation of interacting protofields, their hypothetical "primordial" state without interaction constitutes the causal version of classical "aether", which is quite real physically, but still cannot be perceived experimentally.

The fundamental protofield interaction can be described mathematically by "existence equation", which does not introduce any artificial assumption, but simply fixes the fact of "nonseparable" coupling (interaction) of two entities (protofields) into a single dynamical system, the world. Progressive emergence of the observed world structure is to be explicitly obtained by the unreduced analysis of the existence equation of the following form:

$$\left[ h_\text{e}(q) + V_\text{eg}(q,\xi) + h_\text{g}(\xi) \right] \Psi(q,\xi) = E\Psi(q,\xi), \qquad (1)$$

where $q$ and $\xi$ are practically continuous (but in reality discretely fine-structured), physically real degrees of freedom of the e/m and gravitational protofields, respectively, $h_\text{e}(q)$ and $h_\text{g}(q)$ are the corresponding "generalised Hamiltonians" describing the (unobservable) "free state/dynamics" of protofields without interaction, $V_\text{eg}(q,\xi)$ is the (attractive) interaction potential, $\Psi(q,\xi)$ is the "state-function" describing the (developing) state of the compound system, and $E$ is the generalised Hamiltonian "eigenvalue" in this state (as the following analysis shows, it is always reduced to a measure of dynamic complexity, expressed by the "generalised energy" in the resulting unified description [19]).[*] Note that the state-function $\Psi(q,\xi)$ is *not* the silently introduced wavefunction, but just a function characterising the state of the compound protofield system that will, however, indeed give rise to the causally specified, physically real wavefunction (section 2.2). We also should *not* specify, in our first-principles description, any particular "model" for the fundamental interaction potential and are going to show instead that the world structure emerges for any generic, nontrivial interaction allowing for a "bound", "nonseparable" system state, while particular interaction parameters determine, in a self-adjustable manner, the quantity and quality of emerging perceivable "matter".

The main result of the unreduced, universally nonperturbative interaction analysis of quantum field mechanics [1,2,19-22] is that already such, generic interaction development gives rise to a quite specific, *essentially nonlinear* system behaviour consisting in local, purely dynamical squeeze of an extended protofield portion to a small, "densely packed" volume alternating with the reverse expansion to the previous state. This sequence of "reductions" and "extensions" of the coupled, *a priori* uniform protofields forms an *unceasing* pulsation called *quantum beat* and driven exclusively by the "main", formally homogeneous protofield interaction. It emerges due to the discovered essentially nonlinear dynamics of generic *unreduced* interaction by natural formation of a fractal hierarchy of interaction feedback loops creating the omnipresent *dynamic instability* in such apparently "simple" system. The quantum beat process is none other than the causally specified "elementary particle" (exemplified by the electron), with its squeezed state accounting for "corpuscular" and extended state for "undular" properties of a "quantum" particle, whereas "wave-particle duality" is ensured by permanent dynamic *transformation* of such particle-process between the two states. Because of intrinsic duality in the behaviour of thus obtained elementary particle, we also call it *(elementary) field-particle*. It is important that each reduction/squeeze centre is chosen by the system *at random* among *many* equally possible but incompatible centres, which is described as *dynamic redundance (multivaluedness)* phenomenon.

---

[*]In particular, the "generalised Hamiltonian" and the eigenvalue $E$ in eq. (1) can correspond to the measured localised "coordinate" in a slightly "dissipative" system configuration of the "quantum measurement" type [19,25] (see section 3.1).

All those conceptual novelties are explicitly derived within the rigorous, unreduced interaction analysis (details can be found in refs. [2,19-22,25]) starting from the existence equation, eq. (1), and using the extended, nonperturbative version [23,24] of the well-known optical, or effective, potential method [27,28]. The initial existence equation is thus transformed into the *effective* existence equation depending formally only on the variable ($\xi$) of one of the interacting entities and containing, instead of $V_{eg}(q,\xi)$, the *effective potential (EP)*. It is this latter quantity that contains essentially nonlinear dependence on the eigen-solutions to be found, expressing interaction feedback loops mentioned above. As a result, one obtains *many* equally real, but incompatible solutions of the effective existence equation (and thus the whole problem) called system *realisations*. Being all equally forced to appear by the driving interaction, system realisations should therefore *permanently replace* one another in a *causally random*, *dynamically probabilistic* (or *chaotic*) order, giving rise to natural, first-principles appearance of incompatible, *real* "events" (of realisation emergence) and *a priori* defined *probability* in this unreduced interaction analysis. Usual, invariably perturbative approach of canonical science is equivalent to the cut ("reduction") of all those dynamical links that give rise to the essential nonlinearity and multivaluedness, which leaves one with a single, actually trivial solution-realisation representing as if an effectively *one-dimensional*, artificially averaged *projection* of the real, multivalued system dynamics. It is clear why this severely cut solution of usual theory cannot describe any structure formation and corresponds to only unrealistically "weak", trivial interaction effects. It is equally evident why any perturbative expansion diverges and what would result from the correctly (and universally) integrated series: dynamic multiplicity of permanently changing solutions.

Deprived from any intrinsic source of structure creation, canonical science resorts to its usual, empirically based "tricks" and tries to artificially, formally insert observed but missing structures in its *purely abstract*, "state-vector" or "histories", description, in the form of either direct postulates, or more sophisticated "nonlinear terms" (added to the *postulated* Schrödinger or "density matrix" equation), or "stochastic" influences of ambiguous "environment" (theories of "spontaneous collapse" [29-33], "decoherence" [34-37], or "continuous measurement" [38]), which can only increase, rather than hide, the glaring contradictions of intrinsically deficient single-valued abstraction. The same refers to the whole official "science of complexity" and its occasional applications to micro-world dynamics [39-41]: it uses only empirically based "guesses", or "sketches", or computer-assisted, but always dynamically reduced (perturbative), single-valued and purely abstract *imitations* of natural richness of interaction development results. It is important therefore that our *essential* nonlinearity, leading to the permanent dynamic instability, reduction-extension cycles of the quantum beat process and its intrinsic, dynamic randomness, results simply from the *unreduced* (i. e. truly "rigorous" and "exact") consideration of a generic real, configurationally simple *interaction process*, without any artificial addition of external, mathematical "nonlinearity", "uncertainty", or "stochasticity" always crudely *imitating* and thus essentially *destroying* the unreduced, *naturally emerging* complexity of real processes.

In the considered case of two protofield interaction, compound system "realisation" takes the form of the above squeezed, "corpuscular" state centred around a particular, thus emerging "space point", while other, equally probable realisations are given by other possible centres of protofield reduction/collapse forming other elements of physically real space. In addition to rigorously obtained expressions of the unreduced interaction analysis [2,19-22,25], this result is confirmed by a transparent physical picture of dynamic instability development. Indeed, if an "infinitesimally" small approach of protofields appears occasionally at certain location, then their attraction locally increases, which leads to further approach/squeeze, and so on, until the maximum compressibility value of protofield material is attained in that self-amplifying, avalanche-like process. Such catastrophic, essentially nonlinear dynamics of each quantum beat cycle (concerning both system reduction and extension) distinguishes it from any conventional linear, or even *formally* "nonlinear", oscillation (exploited e.g. in various "vacuum", or "zero-point", oscillation "models" of the universe). It is also physically clear why any real interaction should give a dynamically redundant result: if, for example, $N$ elements of each protofield take part in the interaction process, then the unreduced interaction will provide $N^2$ versions of their combination (or "entanglement") giving rise to $N$-fold redundancy, since the number of "places" in reality for interaction products is the same as for interaction participants. Those "physical" explanations and the whole resulting picture look quite consistent and "generic", which emphasises the triviality of canonical science blunder putting it into the elementary "vicious circle" trap of "self-consistent" perturbation theory and "solution uniqueness theorem" [19] (it is actually *assumed* that the potential is a single-valued function and has a weak effect on the solution, and this assumption is then used to "deduce" that the solution is "indeed unique" and only slightly differs from the "unperturbed" one).*

Because of dynamic multivaluedness phenomenon, the general problem solution for the observed (generalised) system "density", $\rho(q,\xi) = |\Psi(q,\xi)|^2$, takes the form of *dynamically probabilistic* sum of the same measured quantity values for all system realisations:

$$\rho(\xi,Q) = \sum_{r=1}^{N_\Re} {}^\oplus \rho_r(\xi,Q) , \qquad (2)$$

where $\rho_r(q,\xi)$ is system density in $r$-th realisation, $N_\Re$ is the total, dynamically determined number of realisations, and each $r$-th realisation is endowed with the *dynamically derived* probability $\alpha_r$:

$$\alpha_r(N_r) = \frac{N_r}{N_\Re} \quad \left(N_r = 1,...,N_\Re; \sum_r N_r = N_\Re \right), \qquad \sum_r \alpha_r = 1 , \qquad (3)$$

where $N_r$ is the number of "elementary" realisations within the $r$-th observed combined realisation. The sign $\oplus$ in eq. (2) corresponds to the dynamically probabilistic sum of individual realisation densities that has no analogue in usual, dynamically single-valued theory (including official "science of complexity") and means not only that each time one and only one of the densities $\rho_r(q,\xi)$ is "unpredictably" realised (chosen), with the corresponding probability $\alpha_r$, but also that system realisations actually, *permanently replace one another* and this unceasing change is driven by the *same* interaction process that creates each individual realisation structure, $\rho_r(q,\xi)$. The latter quantity is obtained [2,19-22,25] as a well-defined version of *dynamic entanglement* between the interacting degrees of freedom, $q$ and $\xi$, which is a *physically real*, fractally structured, always *transiently* formed (dynamically unstable) and therefore probabilistically changing "mixture" of interacting entities, as opposed to mechanistically simplified, purely formal "quantum entanglement" of abstract "state vectors" discussed in certain "potential" applications of conventional quantum mechanics.

It is important that permanent realisation change implied by the dynamically probabilistic sum of eq. (2) inevitably engenders the existence of intermediate system state through which transitions between realisations can actually proceed. This state is obtained as a special solution of effective existence equation, completing the set of "regular", dynamically localised realisations, and is called the "main", or intermediate, system realisation [19-22]. Before forming each next entanglement of a regular realisation, interaction components (protofields) should disentangle from the previous regular realisation, and this disentangled, quasi-free, but rather irregular, probabilistically changing system state, common for all regular realisations and formed by all transitions between them, just constitutes the intermediate realisation. It provides essential dynamical links between otherwise "independent" system realisations and relates them into a single, complex (multivalued) dynamics. As we shall see in more detail below, the intermediate, dynamically probabilistic realisation provides the *causally* specified, realistic extension of the *quantum-mechanical wavefunction* (section 2.2) universally applicable also at higher levels of complex world dynamics (section 3.2).

---

*The formal demand for wavefunction uniqueness is taken, ironically, as one of the main "postulates" of standard quantum mechanics, contributing thus to its "mysterious" incompleteness.

Individual realisations of the attracting protofield system correspond to dynamically squeezed protofield states, concentrated each around its own "reduction centre" (its position is determined by respective eigenvalues of the effective existence equation), and those centres are chosen in a dynamically random order within quantum beat process of each elementary particle. Since protofield entanglement within squeezed states forms the basis for the physically real space tissue, each particle-process can also be described as a dynamically chaotic, jump-like spatial wandering of its squeezed, "corpuscular" state, or "virtual soliton" (to be distinguished from ordinary solitons, representing dynamically single-valued, and thus essentially linear, solutions of some model equations). Temporal rate of this wandering, i. e. quantum beat frequency, reflects total regularity of reduction-extension event appearance constituting the *real, physical basis of time* in the form of most fundamental clock-work of the world. The intrinsic *spatial* chaoticity of virtual soliton wandering is at the origin of fundamental and universally defined property of *mass-energy (inertia)* of the field-particle, though its quantitative expression, the value of mass-energy, is determined by the temporal rate of this spatially chaotic process.

Different elementary field-particles appear as multiple realisations of the interacting protofield EP determining the values of "intrinsic" particle properties, such as mass. It is related to self-consistent dynamic formation of EP well occurring *together* with protofield reduction in the same process of self-amplified nonlinear squeeze described above. It is important that the totality of quantum beat processes for all elementary particles of the world form a *single, dynamically unified* (and in particular, temporally *synchronised*) process of protofield interaction, which provides a physically sound, causal explanation for the observed *universality* of "fundamental constants" and particle/interaction properties in the whole universe [1,19-22].

Finally, the key property of *dynamic complexity* can be *universally* defined in our approach as a growing function of the number of system realisations or rate of their change, equal to zero for the *unrealistic* case of only one system realisation. Thus, both $\ln(N_{\Re})$ and mass-energy provide possible (and related) measures of system complexity. Note the fundamental difference between our definition of complexity and certain formally similar definitions of usual, dynamically single-valued theory. It is clear from the above definition that any dynamically single-valued description can only give results of zero dynamic complexity. However, canonical science employs its traditional "empirical" cheating, here in the form of formal counting of *arbitrary* observable entities (structure elements) that seem to have relation to intuitively introduced system complexity. This leads naturally to crude errors in complexity estimates and results in well-known contradictions of official "science of complexity" (see e.g. [19,42]), including the evident absence of consistent and universal definition of the "main" studied quantity, dynamic complexity itself. It is that, purely abstract and basically deficient "science of complexity" that starts being used now in quantum behaviour description [39-41], despite the existing totally consistent and first-principles theory within the dynamic redundance paradigm [1,2,19-25] (which is demonstratively ignored by the adherents of canonical abstraction). It is evident that the only possible way to the genuine, *causally complete* science of complexity *cannot* avoid explicit, *rigorous derivation* of structure elements (we call them realisations) giving rise to complexity and its unified definition, and this leads *inevitably* to the dynamic multivaluedness paradigm discovered and used in our universal science of complexity and quantum field mechanics. An indispensable unifying component of that causally complete description of dynamically complex behaviour is just provided by the emerging, *physically real* and universally applicable wavefunction concept.

## 2.2. Complex-dynamic quantization and causal wavefunction emergence

Dynamic discreteness, or *quantization*, of any real, complex-dynamic interaction process is due simply to its unreduced, *holistic* nature ("everything interacts with everything") [1,2,19-22,24,25], and this internally continuous discreteness (step-like inhomogeneity) of the protofield interaction process is the causal, physical origin of "quanta" and related Planck's constant $h$ [1,19] simply postulated in the canonical quantum theory. Natural discreteness and duality of complex, multivalued development of protofield interaction (section 2.1) solve the "mysteries" of canonical, single-valued scheme and take the integrated form of quantum beat dynamics. Each quantum beat cycle produces a physically real, discrete unit of time (as its period) and physically real, tangible space structure (as its elementary dimension, the length of virtual soliton jump in one cycle). This physics determines the dimensions of action for Planck's constant and reveals the extended role of action as universal (integral) measure of dynamic complexity [1,19-22]. Thus, each cycle of the quantum beat dynamics determining the physical essence of elementary particles and their interactions corresponds to a discrete change (actually decrease) of the system action-complexity, $\mathcal{A}$, equal to $h$: $\Delta \mathcal{A} = -h$.

That natural quantization of elementary field-particle moving with the global velocity $v$ can be expressed mathematically as (see refs. [1,2,19-22] for more details)

$$E = h\nu_0 \sqrt{1 - \frac{v^2}{c^2}} + \frac{h}{\lambda_B} v = h\nu_0 \sqrt{1 - \frac{v^2}{c^2}} + h\nu_B = m_0 c^2 \sqrt{1 - \frac{v^2}{c^2}} + \frac{m_0 v^2}{\sqrt{1 - \frac{v^2}{c^2}}} \;, \tag{4}$$

where the total energy, $E$, is given by the *dynamically discrete* analogue of partial time derivative,

$$E = -\frac{\Delta \mathcal{A}}{\Delta t}\bigg|_{x=\text{const}} = \frac{h}{\tau} = h\nu \;, \tag{5}$$

$h\nu_0 = m_0 c^2 = E_0$ being the rest energy (and $m_0$ the rest mass) of the field-particle, describing the totally irregular virtual soliton wandering in the state of rest; the global motion momentum is determined by the quantized version of partial space derivative,

$$p = \frac{\Delta \mathcal{A}}{\Delta x}\bigg|_{t=\text{const}} = \frac{|\Delta \mathcal{A}|}{\lambda} = \frac{h}{\lambda} \;, \tag{6}$$

while the corresponding field-particle velocity is

$$v = \frac{\Delta x}{\Delta t} \;; \tag{7}$$

$\lambda \equiv \lambda_B = (\Delta x)|_{t=\text{const}} = h/mv$ is the *de Broglie wavelength* (and $\nu_B = v/\lambda_B$ *de Broglie frequency*) causally obtained as *emerging* "quantum of space", a minimum directly measurable (regular) inhomogeneity of elementary quantum field with complexity-energy $E$ ($> E_0$) resulting from its global motion [2,19-22]; $\tau \equiv (\Delta t)|_{x=\text{const}}$ is the quantum-beat pulsation period measured at a fixed space point (so that $E = h/\tau$), while $\tau_0 = 1/\nu_0 = h/m_0 c^2$ is the quantum beat period at the state of rest. Those rigorously obtained expressions and definitions of quantities like $\tau_0$, $\tau$, and $\lambda$ provide a *mathematically exact* description of the fundamental, causally specified origin of *physically real* space and time [1,2,19-22].

The total field-particle energy partition of eq. (4) reflects the complex-dynamical structure of the quantum beat dynamics of the moving field-particle that contains a regular global motion component (second term in the sum of eq. (4)) dynamically mixed with purely random deviations (first term in (4)) from that global tendency of spatially chaotic (*partially* ordered) virtual soliton wandering. Relative proportions of the two tendencies of the total particle energy, changing with the global motion velocity, determine the *causally extended* effects of *relativity* naturally *unified* now with causal quantization into the complex quantum beat dynamics [2,19-22] and including relativity of intrinsic time flow in the form of *consistently derived* relation between respective quantum beat periods, $\tau$ and $\tau_0$.

However, whereas *h* and the above quantization expression, eq. (4), describe virtual-soliton, corpuscular state jumps in the unified quantum beat dynamics, the latter contains also the equally important "extended" phase of system transitions between those "regular", localised realisations. The system in this phase should first "delocalize" (expand) itself from the last virtual soliton state, and then, before being squeezed again around a new reduction centre, it passes by highly chaotic and common for all "regular" realisations intermediate state, also called "main", or intermediate, realisation [19-22]. It is this particular system state, constituting an integral, indispensable part of internal field-particle dynamics and *explicitly obtained* in the *complete* solution of existence equation (section 2.1), that provides the causal, physically real extension of the canonical wavefunction concept. Now we can express the naturally quantized, dynamically multivalued character of quantum beat process in terms of its extended, wavefunctional phase, forming the dualistic complement to the above quantization expression in terms of localised state (virtual soliton) behaviour.

State-function $\Psi(q,\xi)$ entering the existence equation and expressing a general system state should incorporate both localised, virtual soliton behaviour described by quantized action complexity $\mathcal{A}$ and the extended-phase, wavefunction dynamics. Therefore the state-function of (massive) elementary field-particle can also be expressed by "wave action", $\mathcal{A}_\Psi$, identified with the total dynamic complexity of the quantum beat process [19-22]. It is given by the product of action-complexity $\mathcal{A}$ and the wavefunction $\Psi$, which reflects their dualistic unity and multiplicative structure of complexity:

$$\mathcal{A}_\Psi = \mathcal{A}\Psi . \tag{8}$$

We can say also that in this expression of the total state-function the first component ($\mathcal{A}$) accounts for discrete action-complexity increments during virtual soliton jumps, while the second one ($\Psi$) provides the (spatial) probability distribution function for dynamically redundant realisations of virtual soliton position (or jump "directions"), expressing the same "unified duality", or causally extended, complex-dynamical "complementarity", of the unreduced interaction dynamics.

If we consider system change in one cycle of the quantum beat process, then it is easy to see that the wave action change after a cycle is zero: $\Delta\mathcal{A}_\Psi = 0$. Indeed, if we start with either extended, or dynamically squeezed system state, then after a cycle the system returns to the same kind of state, physically "symmetric" with respect to the starting one. A more fundamental origin of the wave action permanence is provided by the complexity conservation law [19], where the total complexity can change its form from a "generalised potential energy" (of a wavefunctional, distributed state) to a "generalised kinetic energy" (of a "realised", localised system state), but their sum, the total complexity (expressed here by the wave action) remains unchanged (see also section 3.2). The one-cycle change of eq. (8) gives then:

$$\Delta(\mathcal{A}_\Psi) = \mathcal{A}\Delta\Psi + \Psi\Delta\mathcal{A} = 0 , \tag{9a}$$

or

$$\Delta\mathcal{A} = -h\frac{\Delta\Psi}{\Psi} , \tag{9b}$$

since the characteristic action value is *h*. Coefficient *h* in the last equality is additionally multiplied by a numerical constant, $i/2\pi$, which does not change the physical sense of the above quantum-beat duality expression, but accounts for the difference between undular and corpuscular states in wave presentation by complex numbers (it may also be tentatively attributed to additional quarter-of-period phase shift during quantum beat transition from collapse to expansion):

$$\Delta\mathcal{A} = -i\hbar\frac{\Delta\Psi}{\Psi} , \tag{9c}$$

where $\hbar = h/2\pi$. A *causally substantiated* version of the differential form of quantization rules ("Dirac quantization") is then obtained by using the above definitions of momentum, eq. (6), and energy, eq. (5):

$$p = \frac{\Delta \mathcal{A}}{\Delta x} = -\frac{1}{\Psi} i\hbar \frac{\partial \Psi}{\partial x}, \quad p^2 = -\frac{1}{\Psi} \hbar^2 \frac{\partial^2 \Psi}{\partial x^2}, \tag{10}$$

$$E = -\frac{\Delta \mathcal{A}}{\Delta t} = \frac{1}{\Psi} i\hbar \frac{\partial \Psi}{\partial t}, \quad E^2 = -\frac{1}{\Psi} \hbar^2 \frac{\partial^2 \Psi}{\partial t^2}, \tag{11}$$

where the wave presentation of higher powers of $p$ and $E$ properly reproduces the wave nature of $\Psi$ [19-22], and derivatives should be understood, strictly speaking, in their dynamically discrete version (even though their continuous versions can be acceptable for many practical purposes).

We emphasize the unreduced complex-dynamical meaning of these familiar relations now *causally derived* from the first principles, without any inconsistently inserted, contradictory postulates hiding, as we can clearly see now, the ignorance of real, *essentially nonlinear* and *multivalued* dynamics of the underlying *interaction process*. The basic rule of eqs. (9) is the direct expression of dynamically complementary phases of quantum beat dynamics, involving also the extended, complex-dynamical and realistic interpretation of the origin of action, Planck's constant (see also [1]), and the wavefunction. As such natural diversity of multivalued dynamics cannot be consistently described in principle within the basically single-valued approach of conventional science, the standard quantum mechanics invents and postulates a *purely abstract* construction of "operators" as if "acting" on "state vectors" from abstract "space" according to *formal* rules like those of eqs. (10), (11). It is quite clear now that those mathematical "actions" of imaginary "operators" are but simplified imitation of *real events* of structure formation (realisation emergence) of an adequately described *interaction* process, which provides a completely causal, realistic understanding of the canonical quantum "miracles". It is not difficult to understand [19] that the same causally complete picture extends also the "second quantization" construction of the canonical formalism, with its "creation and annihilation operators": all the elementary particles, as well as their "corpuscular" and "undular" states/manifestations, are *really*, dynamically created and destroyed in the unreduced, dynamically multivalued process of protofield interaction and entanglement, as it is specified in the quantum field mechanics. The well-known ambiguities and contradictions of the "first" and "second" quantizations of various "particles" and "fields" are naturally eliminated within this intrinsically unified picture. Causal quantization is also directly related, through eqs. (5)-(6), to the emergence and quantized character of the physically real space and time, so that the "wavefunction quantization" of eqs. (10)-(11) can be considered as expression of (dynamic) quantization of space and time in terms of de Broglie (and Compton) wavelength and quantum beat period [19-22]. The same causal quantization picture involves the extended, realistic interpretation of "(Heisenberg) uncertainty relations" of standard theory, which are *consistently derived* now from quantum beat dynamics of the protofield interaction process (see eqs. (5), (6)) and, being another manifestation of its complex-dynamical discreteness, refer to *both* "corpuscular" and "undular" states of the system [19-22].

The main dynamic equation for the wavefunction, or "wave equation", can now be obtained by insertion of causal quantization expressions, eqs. (10)-(11), relating the "localised", corpuscular manifestations of quantum beat dynamics to its "undulatory" parts, into dynamic quantization expression in terms of corpuscular aspects, such as eq. (4) (note the difference between this logically transparent, "algebraic" substitution and obscure "operator substitution" of standard quantum mechanics as if "magically" transforming "ordinary", classical-world relations into "mysterious", "quantum" equations). Being applied to the full, relativistic version of eq. (4), this step leads to the simplest forms of Klein-Gordon and Dirac equations that can then be extended to include e/m and gravitational interactions [19]. In this paper we are interested rather in causal derivation of ordinary, nonrelativistic Schrödinger equation and therefore shall insert quantization rules into the nonrelativistic limit of eq. (4), where the rest energy is subtracted and interaction potential is added,

$$E = \frac{p^2}{2m_0} + V(x,t), \tag{12}$$

the "potential" (energy of interaction) $V(x,t)$ expressing the "latent", "potential" part of the total complexity-energy due to particle interaction with another, unspecified system (generally composed of other elementary particles), which gives rise to the next, higher (sub)level of complexity [19] (therefore $V(x,t)$ in eq. (12) should not be confused with the total underlying *protofield* interaction potential of eq. (1) acting at the lowest level of complexity and *eventually* giving rise to $V(x,t)$, after elementary particle formation at that lowest level, represented by the rest mass $m_0$ in eq. (12)). This *potential* form of complexity, or "dynamic information", is to be transformed into its "unfolded", *explicit* form of "generalised kinetic energy", or "dynamic entropy", so that the sum of both forms, the *total complexity*, remains unchanged [19], while eqs. (4), (12), and resulting wave equations emerge as manifestations of this *universal law of conservation and transformation of complexity* (cf. section 3.2).

Substituting eqs. (10), (11) into eq. (12), we get the "time-dependent" form of Schrödinger equation:

$$i\hbar \frac{\partial \Psi}{\partial t} = -\frac{\hbar^2}{2m_0}\frac{\partial^2 \Psi}{\partial x^2} + V(x,t)\Psi(x,t) \ , \tag{13}$$

which is easily transformed into a time-independent form for a potential, $V(x)$, that does not contain externally imposed time dependence:

$$-\frac{\hbar^2}{2m_0}\frac{\partial^2 \Psi}{\partial x^2} + V(x)\Psi(x) = E\Psi(x) \ , \tag{14}$$

We have thus consistently *deduced* the Schrödinger equation for the *causally extended* wavefunction, without using any "quantum" or "classical" postulates (note that $x$ here can be vector particle position in a three-dimensional space and many-particle "string" from a causally extended "configuration space", see also below). The novelty of the causally substantiated equation is not in appearance of extra terms, which would be ambiguous in view of extensively verified precision of usual version, but in the profound, *causally complete* understanding of the underlying complex-dynamical, essentially nonlinear interaction process, which provides a clear, demystified explanation for canonical "quantum postulates" and related "rules". One of these rules, "Born's probability postulate", revealing the *intrinsically probabilistic* meaning of the wavefunction, is naturally incorporated, as we have seen, in the *dynamically multivalued* origin of the causal wavefunction, but it can also be obtained in a more formal way, within the "dynamical boundary/initial conditions" of the unreduced complex dynamics [19-22] (dynamical "matching" of "regular" realisations to the wavefunction which they directly produce in the "intermediate" realisation phase). The underlying complex dynamics is described by the extended EP formalism, which is only briefly mentioned here (section 2.1), and provides the "double solution with chaos" [2,19-22] that has more directly observable effects and consequences rather in relativistic quantum mechanics (like chaotic virtual soliton wandering providing the causal origin of mass, etc.) and at higher sublevels of complexity obtained from the same unreduced analysis of eqs. (13), (14) (see also section 3) [19,23-25].

In order to better demonstrate the complex-dynamical origin of "ordinary" Schrödinger equation, we multiply eq. (14) by $(m_0/\hbar^2)\Psi^*(x)$ and integrate over the domain of $\Psi(x)$, which gives

$$q^2 + \frac{m_0}{h}\frac{V_\Psi}{h} = \frac{m_0}{h}\frac{E}{h} \ , \tag{15}$$

where

$$V_\Psi = \int_\Omega \Psi^*(x)V(x)\Psi(x)dx \ , \tag{16}$$

and

$$q^2 = -\frac{1}{8\pi^2}\int_\Omega \Psi^*(x)\frac{\partial^2 \Psi}{\partial x^2}dx = \frac{m_0}{h}\frac{K}{h} \ , \tag{17}$$

with $K$ representing "kinetic energy" (we use an estimate, $\Psi(x) \sim \Psi_0 \exp(ikx)$, in the integral below):

$$K = -\frac{\hbar^2}{2m_0} \int_\Omega \Psi^*(x) \frac{\partial^2 \Psi}{\partial x^2} dx \sim \frac{\hbar^2 k^2}{2m_0} = \frac{p^2}{2m_0} \ . \tag{18}$$

Taking into account the complex-dynamical interpretation of the property of mass/energy [19-22] as being due to unceasing chaotic sequence of quantum jumps of the virtual soliton, each of them involving the action-complexity change of $h$, we can see now that the Schrödinger equation, eq. (15), corresponds to the complexity conservation law for the first two sublevels of world complexity, expressed in terms of elementary complexity quantum, $h$. Every number of such quanta within both sublevels is obtained as a product of the internal rest-mass complexity, $m_0/h$, of the lowest sublevel of dynamics by the respective part of higher-sublevel complexity ( $E/h$, $V_\Psi/h$, or $K/h$ ).

    This interpretation provides a causally complete explanation for the occurrence and sense of characteristic, "quanta-bringing" coefficient, $\hbar^2/m_0$, in the Schrödinger equation, eq. (14). It also leads to the consistent, physically complete understanding of the origin of *energy level discreteness* of quantum particle in a binding potential well (the famous "energy quantization" [1]), the phenomenon that is only *mathematically* deduced in the conventional Schrödinger formalism [6] *postulating* its essential points and thus actually *hiding* within them the ultimate, physical origin of "quantization". We can see now that each discrete energy level is determined by the discrete, integer, number of physical, causally emerging "quanta of complexity" it contains, in the form of integer number of quantum-beat pulsation cycles confined within "rigid" potential walls (see ref. [19] for more details). The physical origin of confined "particle wave" is the same as that of de Broglie wave of isolated (free) elementary particle (see eq. (4) above and refs. [2,19-22]): it is the average, global-motion tendency in the chaotic quantum-beat wave field. The difference with respect to the "standing-wave" explanation of the standard theory is that the *abstract* and therefore inconsistent, linear "wave" of the latter is now completed by the essentially nonlinear, dynamically multivalued (and therefore spatially chaotic) quantum beat pulsation hidden within that formally linear "envelope", but providing the wavefunction and observed probabilistic *particle emergence* with the totally realistic, causally complete meaning. In other words, every conventional, linear or "nonlinear", "oscillation" of a quantum particle in a well, described by the Schrödinger equation, is in reality synchronised with, and actually governed by, the underlying more fundamental quantum beat pulsation of the interacting protofield system (or rather its particular realisation for the field-particle in question), which explains the appearance and universality of $h$ in quantum system dynamics (see also ref. [1]).

    The standard quantum mechanics, as well as all its officially accepted, "formal" or "causal", modifications actually bury the multivalued dynamic reduction of the underlying, essentially nonlinear interaction process under the formally imposed postulate of wavefunction single-valuedness, which provides usual theory with deceptive simplicity of a linear framework, but strictly prevents any causal understanding with evidently "nonlinear", highly inhomogeneous and intrinsically probabilistic events of the "true", localised particle emergence in the same, undular system dynamics (cf. famous Einstein questions about real particle appearance from its wave at the Fifth Solvay Congress [43]). Conventional quantum mechanics, and actually all canonical science as well, gets in that way into the elementary "vicious-circle" trap of false "self-consistency" inherent in the dynamically single-valued approach and most clearly seen in the disruptive "logic" of canonical "uniqueness theorems" and perturbation theory: once uniqueness of system dynamics (including the values of interaction potential, solution, etc.) is explicitly or implicitly inserted as a starting *assumption*, one *cannot* really confirm it by a complete solution, but can only restate it as a result of *reduced* solution [19]. Being however accepted, that "small lie", or logical "trick", gives rise to a whole series of "magic" facts taking the form of "inexplicable" quantum "paradoxes" fixed by "quantum postulates" and developing into the practical absence of *any* rigorous, unified science at *higher* complexity levels, where the number of necessary "postulates" quickly diverges to infinity.

It would be not out of place to conclude this brief description of the causally extended, complex-dynamical concept of quantum-mechanical wavefunction by emphasising the way in which that extended theory resolves the well-known basic contradictions of usual wavefunction. One group of traditional wavefunction difficulties is related to its postulated *abstract-probabilistic*, rather than realistic (causally deduced here) interpretation. We have seen above how dynamic redundancy of unreduced interaction process provides the ultimate, causally complete *origin of randomness* and dynamic, *a priori* defined probability in quantum, and actually any other [19], dynamics, which cannot be revealed in principle within the dynamically single-valued, or *unitary*, approach of standard theory. It is the dynamically multivalued, permanently and chaotically changing wave field of causally extended, real wavefunction of interacting protofield system that explains the "inexplicable" probabilistic properties of the canonical wavefunction. In particular, the probabilistic "Born postulate" and its "normalisation" procedure, as if emphasising the unreal wavefunction character in standard quantum mechanics, is explained as a natural, inherent property of complex-dynamical "intermediate realisation" that, being the unifying (and physically real) transitional state of the protofields between their "regular" (explicitly observable) states/realisations, indeed permanently and "automatically" adapts to those "eigenstates" which, in their turn, dynamically *emerge* in accord with external limiting factors (measurement, interaction potential, etc.). This is but a particular manifestation of the universal property of *dynamic adaptability* of complex (dynamically multivalued) systems [19].

A related causal solution is obtained for the *"configurational" difficulty* of the wavefunction for a many-particle system, mentioned already in the original Schrödinger theory [9]: the wavefunction of an *N*-particle system depends on each of "independent", and irreducible, coordinates of "coherent" particles forming the system and thus on system "configuration" in abstract, 3*N*-dimensional, rather than real, 3-dimensional space. In the extended, complex-dynamical description of the same situation we have physically real, *explicitly obtained* emergence of system configurations as multiple, permanently changing realisations in *ordinary, three-dimensional space*. That ordinary physical space also represents the simplest possible system "configuration", which is silently postulated as "natural" and "realistic" in scholar theory, but is now causally *derived* by the unreduced protofield interaction analysis explaining, in particular, the exact number of three of those "basic" spatial dimensions [1,19-22]. Larger numbers of *effective* "dimensions" emerge in the complex-dynamic approach in the same way and correspond to growing realisation numbers at higher complexity levels of a naturally developing interaction hierarchy, giving rise to respective levels of tangible "configuration" space. Real, three-dimensional particles and their groups permanently and probabilistically *jump* between those multiple realisations/configurations at their corresponding levels, providing thus "impossible" *unification* of their reality *and* multiplicity (by contrast, they should necessarily be classified into *abstract* "dimensions" of abstract "configuration spaces" within the dynamically *single-valued* description of standard theory). The unreduced, dynamically multivalued picture, rigorously described by the above universal equations (eqs. (1)-(3), see also section 3.2 below), specifies also the causally complete, *dynamic* origin of *"quantum indistinguishability"* of particles of the same species and "Pauli's exclusion principle" (two or more squeezed, virtual-soliton states cannot coexist within one "global" route of their jumps) [19].

The same, unified explanation includes causal interpretation of *"quantum linear superposition"* of possible states, among which, however, only one can appear in actual measurement. This another "inexplicable" rule of standard quantum mechanics questioned by Schrödinger in his famous "gedanken" experiment with a cat [15] ("Schrödinger's cat" paradox) is a direct expression of complex, *multivalued* system dynamics: the real quantum system described by "enveloping" linear superposition of states actually performs *permanent*, probabilistic *jumps* between all component "eigenstates"/realisations, with its "transitional" state just being described by the (now causally extended) wavefunction, whereas the whole jump-like behaviour is due to the underlying dynamically nonlinear, multivalued interaction process(es). When such "compound" state undergoes explicit "quantum measurement" — which is but generic higher-level interaction with small dissipativity [19,25] (see section 3.1) — it means that the system is occasionally, but irreversibly "caught" by that higher-level interaction at the moment it takes one of intrinsic "eigenstates" (realisations) and then this event is "macroscopically" amplified by the "measuring device" due to dissipativity (openness) and the same *intrinsic instability* of unreduced interaction dynamics (now at its higher levels). However, starting already from the beginning of (microscopic) measurement/interaction process, the measured system cannot continue its chaotic jumps between eigenstates and remains for a long enough, though also only transient, period of time in that "snapshot" eigenstate (or their dense group). We emphasize the fundamental difference of this causally complete, complex-dynamic hierarchy of real interaction processes from technically over-sophisticated, but senseless speculations of unitary theory about "environment-induced decoherence/collapse" [29-38], "Schrödinger-cat states" (e. g. [44]), etc. In particular, a macroscopic/semi-classical or many-particle *quantum* state is characterised by exactly the same kind of physically real, dynamically chaotic "quantum jumps" between its "eigenstates"/realisations, or "configurations" (they can form several sublevels), as the most elementary quantum system (isolated elementary particle).

The same kind of completely specified, essentially nonlinear and dynamically redundant process constitutes the true physical origin of *"quantum entanglement"* of states, related "paradoxes" and "quantum miracles" involving the wavefunction (e.g. "quantum teleportation"), whereas the "officially accepted" science, proud of its "rigour" and thoroughly excluding alternative approaches from any its printed source, sinks deeper and deeper into the ever growing swamp of crudely simplified (linear) but technically "entangled" and heavily mystified abstractions, ambiguous "experiments" and related "post-modern", verbal exercises around the same, invariably reduced and ruptured image of esoteric, "veiled" quantum reality. We can see now that the main, essentially nonlinear part of physically real, multivalued quantum dynamics is indeed "fundamentally hidden", but only from the trivially simplified, dynamically single-valued approach of conventional, "symbolical" science. On the other hand, we can specify the exact dynamic origin of *apparent linearity* (or "unitarity") of Schrödinger wavefunction evolution: causal wavefunction describes the *transiently* free, indeed effectively linear (weak-interaction) system state, which is permanently "punctuated", however, by *explicitly nonlinear* (strong-interaction) states of "virtual soliton" (as it is reflected by the inevitably "contradictory" postulates of standard theory).

The proposed complex-dynamical explanation of canonical "quantum miracles" is indirectly supported also by its universal applicability at all higher complexity levels [19,24,25], providing the intrinsically unified picture of *physically unified* world dynamics devoid now of "unsolvable" problems and glaring inconsistencies so characteristic of canonical, dynamically single-valued science (cf. section 3.2). Particular features of "quantum" complexity sublevels forming their specific, "mysterious" aura are due to their unique, *lowest* position in the universal hierarchy of complexity [19], so that many details of the unreduced, complex-dynamical version of quantum behaviour, such as "quantum jumps" between redundant realisations, can be observed only as a whole, by their "final" results rather than detailed structure, which gives rise to traditional mystification. It is important that the latter cannot be eliminated by the official, unitary "science of complexity" (cf. [42]), which uses the same, dynamically single-valued, basically limited approach (see [19] for more details), even though it readily produces empirical, computer, verbal, philosophical, and abstract *imitations* of complexity, including e. g. *empirically* fixed multiplicity of necessarily *coexisting* motion states ("unstable periodic trajectories", or "attractors") in *abstract* "phase spaces" [39-41].

It is important to see the difference between the proposed causally complete, first-principles and dynamically nonlinear (multivalued) picture of quantum behaviour and a growing number of imitations of realistic/causal description actually obtained within the same unitary science paradigm, but using "causal" terminology, or "interpretation", and often representing a mechanically reduced deformation of the truly realistic picture of complex-dynamical wave mechanics. We can only briefly refer here to the so-called "Bohmian mechanics" [45-48], first produced by Louis de Broglie [16,43] as a *deliberately* reduced version of his complete, "double solution" theory [17,18], then "rediscovered" by Bohm [45] and now extensively propagated under the name of "(causal) de Broglie-Bohm interpretation", without reference to the unreduced double solution of de Broglie (see also [2]). Another group of imitations is centred around so-called "statistical electrodynamics" and "zero-point field theory" that *postulates* the existence of ambiguous, abstract "vacuum fluctuations" and tries to "deduce the world" from their *essentially linear* dynamics (see e. g. [49-52]). Recently this approach has "enriched" itself with another simplified imitation of the level of "de Broglie-Bohm interpretation" [53] that looks very much like mechanically reduced adaptation of certain ideas of the quantum field mechanics [19-21]. That sort of silent "borrowing" and simultaneous crude, mechanistic simplification of original realistic theories seems to be a favourite "method" of the false-causality approach initiated yet at the beginning of "new physics" by its most "prodigious" stars. Without going into details of mentioned and numerous other (e. g. [54]) versions of mechanistic "causality", note their key, fundamental distinctions from the unreduced version of quantum field mechanics, which they totally divide with the standard quantum theory: they *postulate* purely *abstract* images of "physical" entities and their properties (like "particles" and "transporting wave" in the Bohmian mechanics, or "physical vacuum" and its "fluctuations" in the "zero-point field" imitations), without any consistent explanation of their origin; they take for granted and use key equations, and thus postulates, of standard, often formally "denied" theories (thus the "causal" Bohmian mechanics simply *rewrites* the same Schrödinger equation for the explicitly separated real and imaginary parts of the wavefunction, instead of its first-principles derivation); in their specific "calculations", they remain totally within the basically linear, perturbative, dynamically single-valued analysis of unitary science; and as a result, the unreduced dynamic complexity of their abstract constructions, measured by the number of *explicitly obtained* realisations, is strictly zero, which is a clear and exact demonstration of their *conceptual equivalence* to the canonical, totally abstract science that does not pretend at least for any special "realism" and "causality". The false causality adherents want thus to obtain too much for too little and do not hesitate to use explicit, unambiguous deviations from elementary scientific (and general) honesty. A particular manifestation of the basic deficiency of "causal" imitations of complexity is the absence within them of any truly realistic and consistent interpretation of the wavefunction, with all its "contradictory" properties, which is not surprising in view of the central, unifying role of this entity in the unreduced complex dynamics.

### 3. Dynamically emerging hierarchy of world complexity and the universal Schrödinger equation

#### 3.1. General solution of Schrödinger equation: True quantum chaos and causal quantum measurement

We have demonstrated, in the previous section, how the causally extended wavefunction and Schrödinger equation it obeys naturally emerge from the unreduced analysis of interaction process in a simple, *a priori* homogeneous system of two coupled protofields. The irreducibly complex, multivalued system dynamics, "hidden" behind the externally linear envelope of canonical formalism, involves explicit *creation* of new entities in the protofield interaction process, emerging as spatially chaotic quantum beat processes, or "elementary particles". Therefore the causally extended wavefunction, Schrödinger equation, and elementary world structures they describe emerge *all together* in quantum field mechanics, which means that the unreduced analysis of the universal science of complexity provides the *exact*, totally complete picture of appearing objects and interactions, in their full "temporal" (cosmological, or "vertical") continuity and "spatial" (structural, or "horizontal") unity. Unrestricted universality of the basic mechanism of system splitting into redundant realisations of dynamically entangled interaction components allows for the natural reproduction of the same phenomenon at each next, dynamically emerging level of the hierarchy of complexity, in both reality and its exact representation by the dynamic redundance paradigm: (grouped) realisations of the first level of complexity, forming the simplest material entities of the world (elementary field-particles), interact (entangle) among them and give rise to the same kind of *dynamically multivalued* structures of the next higher level of complexity (e.g. atoms), and so on, up to the most sophisticated products of real world dynamics usually described only within fundamentally inexact, "humanitarian" disciplines [19].

That *intrinsic creativity* of unreduced interaction processes, totally absent in the basically perturbative description of canonical, dynamically single-valued theory, gives rise to the "magic" status of any explicitly complex, dynamically creative effects of permanent realisation emergence and disappearance, so common in quantum behaviour and causally explained within the complex-dynamical extension of usual Schrödinger formalism (section 2). The created complex-dynamical objects of the first complexity level, elementary field-particles, start interacting among them by exchanging *physically real* but massless (quasi-regular) perturbations, photons, initiated by quantum beat processes within particles, but becoming perturbations of another, weak-interaction kind outside of them [19]. Those interaction processes of the second (sub)level of complex world dynamics are just described, in the nonrelativistic limit, by the Schrödinger equation, eqs. (13)-(14), now causally derived, without any "mysterious" postulates, from the unreduced dynamics of the first level of protofield interaction development.

It follows that this second level of protofield interaction development gives rise to the second complexity level described by the unreduced (multivalued) solution of Schrödinger equation. Indeed, if we put $x = (q,\xi)$ in eq. (14), where $q$ accounts for the degrees of freedom of a "separable", effectively one-dimensional problem part, and $\xi$ describes all other (generalised) coordinates making interaction "nonseparable", then the same Schrödinger equation can be written as

$$-\frac{h^2}{2m_0}\left(\frac{\partial^2 \Psi}{\partial q^2} + \frac{\partial^2 \Psi}{\partial \xi^2}\right) + \left[V_0(q) + V(q,\xi)\right]\Psi(q,\xi) = E\Psi(q,\xi), \quad (19)$$

where $V_0(q)$ is the effectively one-dimensional (separable) part of the total interaction potential (obtained, for example, by averaging over $\xi$). We obtain a particular case of our universal existence equation, eq. (1), where now

$$h_e(q) = -\frac{h^2}{2m_0}\frac{\partial^2}{\partial q^2} + V_0(q), \quad (20)$$

$$h_g(\xi) = -\frac{h^2}{2m_0}\frac{\partial^2}{\partial \xi^2}, \quad V_{eg}(q,\xi) = V(q,\xi), \quad (21)$$

We can apply again the results of unreduced existence equation solution, in the form of dynamically redundant, probabilistically changing system realisations [24,25] (section 2.1) and write the *general solution* of the Schrödinger equation for interacting particle(s), eq. (19), in the form of probabilistic sum of eq. (2), which refers to the observed system density, $\rho(q,\xi) = |\Psi(q,\xi)|^2$, and implies the *dynamically determined* probability of each realisation emergence, eq. (3), as well as the intermediate realisation, or causal *wavefunction* (or generalised *distribution function*), now of this *new*, second level of complexity (we refer to other works [2,19-25] for detailed expressions for the component realisations densities, $\rho_r(q,\xi)$, obtained within the generalised EP formalism). According to the causally extended "Born probability rule" (section 2.1), that wavefunction is defined on the *physically real* space of dynamically redundant configurations ($\equiv$ realisations) of interacting particle system described by the Schrödinger equation, eq. (19), and provides the distribution (density) of configuration/realisation probabilities.

We have here a particular case of complex-dynamical emergence of physically real "configuration space" described at the end of previous section and showing that "space" *in general* is none other than dynamically emerging *hierarchy* of those causally obtained realisation spaces of respective complexity levels, where the most fundamental (and most fine-grained), "embedding" physical space (the direct complex-dynamic extension of "Newtonian" space) is simply obtained as the very *first* level of world complexity, corresponding to elementary field-particles and their characteristic dimensions, such as the Compton wavelength [19-22] (section 2). The same hierarchy of complexity levels, limited from below, determines the structure of time specified, at each level, as the frequency of realisation emergence events (appearing in a *dynamically random order*), with the quantum beat frequency corresponding to the most fundamental (and most fine-structured/precise) time flow of the world.

The spatial and temporal aspects of each level of complexity are dynamically unified by the causal, dynamically probabilistic wavefunction, or *distribution function*, of each of the corresponding systems represented by its intermediate realisation, which gives rise to the hierarchy of wavefunctions and corresponding Schrödinger equations (see section 3.2). Note that the complex-dynamical wavefunctions at the quantum sublevels of complexity can be considered also as causal extensions of the corresponding "density matrices", the latter being a semi-empirical, postulated, and therefore contradictory imitation of the unreduced wavefunction within the standard, dynamically single-valued quantum mechanics. The (extended) "density matrix" of the first level of complexity coincides with the (extended) Schrödinger wavefunction, and at that, lowest level of complexity the appearing contradictions of the conventional version can be at least correctly fixed with the help of a limited number of "quantum postulates" (remaining, however, provocatively "puzzling"). The higher sublevels of quantum complexity involve a growing diversity of complexity manifestations, and the contradictions of the corresponding single-valued projection of the "density matrix formalism" cannot be so easily "compensated" by "postulates", which leads to the directly incorrect results of density matrix applications. The dynamically random structure of the emerging system configurations that should be consistently *deduced* in the *purely dynamic* analysis (as it is given by the universally applicable expressions of eqs. (2), (3)) is artificially *inserted* by implicit assumptions of the canonical density matrix formalism that "confirms" them by its results, which creates a characteristic "vicious circle" of the dynamically single-valued reduction of reality (cf. section 2.2).

Returning to the unreduced interaction of quantum particles, note that the obtained general solution for the corresponding Schrödinger equations and the method for its detailed derivation are applicable to various configurations of the system of quantum particles with interaction. The key difference of this truly complete general solution from any "general solution" of the unitary science is the *dynamically probabilistic* summation of *many* realisations, *each* of them resembling such ordinary "general solution", where the *dynamically* derived realisation *probabilities* play the role of complex-dynamical "expansion coefficients". This comparison demonstrates also the difference between the complex-dynamical and unitary ideas about "completeness" of solutions. It is not difficult to understand that the truly dynamical, truly random (probabilistic), and unceasing realisation change at the level of interacting quantum particles provides the causally extended, noncontradictory version of "true" *quantum chaos* [19,23-25], as opposed to its conventional, dynamically single-valued imitation [55-58] that cannot provide the intrinsic, purely dynamic origin of randomness, being the main property of any "chaos", and is obliged therefore to substitute it for a "very entangled" (but basically regular) behaviour and/or an artificially inserted randomness of "external influences" (one has here another example of the basically incorrect application of the conventional approach of the "density matrix" type).

The unreduced dynamical chaos in quantum systems can have two qualitatively different manifestations. If one has the effectively "nondissipative"/closed, or "conservative" system configuration (usually with a small number of interaction participants), then one deals with the case of *Hamiltonian quantum chaos*, actually constituting the only case considered in the canonical quantum chaos theory. In this case the configurations of incompatible system realisations that permanently and randomly replace each other are of the same general "dimension" as the "projected", single realisation of its spatially extended, undular ("wavefunctional") state obtained within a conventional, perturbative approximation, so that the real system indeed chaotically and permanently changes its configuration, but the latter remains generally delocalised and qualitatively similar to what one would expect from any standard Schrödinger wave dynamics. Usually one deals in such cases with a conservative system interacting with a spatially or temporally periodic external "perturbation", and the unreduced, dynamically multivalued solution of the corresponding Schrödinger equation (like eq. (19)) leads to prediction of more, or less, pronounced, truly chaotic behaviour, in the form of the probabilistically changing system realisations (eq. (2)), that correctly passes to the corresponding classical system behaviour after the ordinary quasi-classical transition [19,24], contrary to the fundamental difficulties with the "correspondence principle" arising in the conventional, dynamically regular and purely abstract quantum chaos theory [59].

If one deals with a slightly dissipative/open system usually containing locally excitable objects (like atomic electrons), then using the same, *dynamically multivalued* solution of Schrödinger equation, eq. (19), or other form of existence equation, eq. (1), one can obtain the causally complete description of a process known as *quantum measurement* [19,25]. The difference with respect to Hamiltonian quantum chaos is that here redundant realisations have strongly localised configurations determined by excited particle positions, and therefore by taking each of those realisations (with the corresponding, dynamically determined probability, eq. (3)), the system is *transiently* "reduced", i. e. dynamically squeezed, to the corresponding location, starting from its extended-wave configuration. After that the system usually reconstitutes its extended-wave configuration (since one has a relatively *small*, spatially confined dissipativity), and continues its motion in the regime of Hamiltonian quantum chaos (or in a free-motion state). In this respect the quantum measurement situation resembles the reduction-extension dynamics of free elementary particles at the lowest level of complexity, but at this higher sublevel we have only one (or just a few) "quantum beat" cycles. Note that fundamental quantum beat processes of participating elementary particles always continue *within* both quantum measurement and (Hamiltonian) quantum chaos dynamics, so that the "extended" system wavefunction contains local component beats within it (they are necessarily, though only partially, synchronised with respect to each other, in both time and space, for a "coherent", "essentially quantum" system [19]). Moreover, usually quantum measurement processes constitute a complexity sublevel situated between those of fundamental quantum beat and Hamiltonian quantum chaos, so that we have in general an almost conservative quantum system performing chaotic transitions between its extended wave realisations, where each realisation contains occasional, more rarely taken realisations of local reduction-extension events of quantum measurement type, and both "extended" ("elastically scattered" or "freely moving") and "localised" ("measured") system states contain unceasing cycles of very frequent (and strong) reduction-extension events of dynamically synchronised quantum beat processes. The difference between Hamiltonian and quantum measurement cases of general quantum chaos clearly demonstrates also the purely dynamic origin of randomness (multivalued dynamic instability) in both cases, since the influence of dissipativity in the quantum measurement case leads simply to change of resulting *configuration* type of ever *dynamically redundant* realisations. By contrast, the official, *dynamically single-valued* theory of quantum (and classical) chaos tries to *artificially reconstitute* the basically absent dynamic origin of randomness and mechanically *inserts* its *total* randomness in the form of *postulated* "noisy" influence of ambiguous "environment" [60-62] *regularly* "modified" (usually "amplified") by the system (standard "exponential amplification" of perturbations is in itself a *basically incorrect* concept of scholar chaos theory resulting from arbitrary, inconsistent extension of perturbative description beyond its validity range [19]).

In that way the causally complete description of quantum system interaction processes reveals the true, realistic and irreducibly complex-dynamical, content of the famous quantum measurement phenomenon [25], resolving all canonical "mysteries" as being due to the basically deficient, dynamically single-valued (unitary) approach of usual theory, and shows quantum measurement dynamics as an *integral part* (sublevel) of multilevel complex dynamics of a system with interaction having a sufficiently low, "quantum" value of its total complexity. If interaction within the system is strong enough to transform the sublevel of transient interaction during measured system reduction to the superior level of *permanently bound system* (like atom), then one deals with the simplest case of *classical* (permanently localised, or "trajectorial") behaviour forming thus the next higher complexity level. This qualitative transition can be explained within the same, *purely dynamic* picture of *unreduced* interaction process as being due to the *independently* random character of quantum beat processes of bound system components, which leaves very little probability for occasional series of quantum jumps of system components in the *same* direction [1,19-22]. This result devaluates completely any "decoherence" theories of classicality emergence within conventional "interpretations" [34-37] relating its very origin to ambiguous "decohering influences" of ever changing "environment" (even though the latter can indeed influence *quantitative* transition parameters) and often falling into the same characteristic trap of unitary "self-consistency" (section 2.2) where what should be obtained is silently inserted from the beginning and then "confirmed" by the "results" (incorrect "density matrix" use, etc.). Indeed, wave coherence destruction by a generally weak "influence" should not necessarily lead to *qualitative* transformation of the very system type from the linearly superposable, undular "wavefunction" to a permanently localised, highly nonlinear "body", indispensable for the emergence of truly "classical", trajectorial behaviour.

Unreduced involvement of hierarchically organised levels of real system complexity provides a *reasonable*, adequate price for the definite solution of the "great mystery" of usual quantum mechanics, since the obtained realistic picture is a *natural* and therefore *qualitatively simple*, but *conceptually*, dynamically *nontrivial* one, especially with respect to artificial limitations and technical sophistication of the dominating single-valued approach (one can see also the well-specified *fundamental* origin of "strange" weakness of the "omnipotent" canonical science in face of "irresolvable quantum mysteries" that has been comfortably hidden in the very basis of single-valued scholar *Weltbild*). It is important that all that self-developing involvement of universal interaction hierarchy is obtained as a sequence of *unreduced solutions* of the main dynamic equations, where the Schrödinger equation, as we have explicitly shown, constitutes both the most important result and universally extendible source of unlimited complexity development process (see also the next section). Therefore the proposed *dynamically multivalued solution* of "ordinary" Schrödinger equation (eqs. (13), (14), (19)) and its applications [19,23-25] are *as much important* for its extended understanding as its causally complete derivation (section 2). One actually obtains a hierarchy of dynamically unified, generalised Schrödinger equations, together with respective causal wavefunctions and probabilistic realisation sets, at all progressively emerging levels of dynamic complexity (section 3.2). Each next (sub)level of complexity is provided with the causally complete version of (generalised) Schrödinger equation "coming" from the neighbouring lower complexity level and having dynamically redundant solutions (realisations) that form real structures of this and higher sublevels, unified in one complex-dynamical process by *causally extended* "density matrix", or wavefunction, or distribution function, explicitly obtained as intermediate system realisation [19]. This procedure of the universal science of complexity eliminates the basic inconsistency of usual approach that tries to formally impose (postulate) a "density matrix" or a "modified Schrödinger formalism" at a given level of dynamics and then uses it to "derive" system behaviour at the *same* level, falling thus into the "vicious-circle trap" of unitary paradigm (section 2.2).

Note also the basic difference of the explicitly derived phenomenon of dynamic multivaluedness (and the resulting "dynamic redundance paradigm") from various formally imposed, non-dynamical hypotheses of "many-worlds" type becoming popular in the "interpretational" part of official quantum theory. Whereas ambiguous "other worlds" serve as the "last resort" for the canonical approach trying to attenuate the evident contradiction between creative misery of its single-valued, unitary evolution and natural richness of Being already at its "quantum" levels (multiple, inter-changing states and species of particles and fields), causal randomness of dynamically redundant system emerges just *due to the real world uniqueness*,* where every system, instead of choosing one, *fixed* configuration from their *global, postulated* multiplicity (that *needs* therefore to be fully realised "elsewhere"), *permanently* and *locally changes* its configuration, with all its observed realisations being *explicitly obtained* from the essentially nonlinear development of the "main", generic and configurationally "simple", interaction of the system. The difference arises simply due to the really complete, universally nonperturbative interaction analysis in the universal science of complexity containing no technical/logical "tricks" or artificial "principles" (in contrast to numerous imitations of "complexity" and "causality" in quantum mechanics and beyond).

Contrary to emergence of the very first complexity level (elementary particles) that can be "experimentally confirmed" rather by its main results (see [1]) than detailed development, multivalued dynamics of next "quantum" complexity sublevels provides a much larger scope of possible applications. Indeed, it follows from the above that practically *any* real interaction of quantum systems, starting from elementary particles, gives rise to the true dynamic randomness, in the form of either genuine quantum

---

*Apart from a general possibility of existence of *quite* different, dynamically "remote" universes, which are *not* "quantum copies" of our universe and can be only weakly connected to it at certain locations.

chaos, or quantum measurement, or their combination. This conclusion leads to the idea of qualitatively extended, "chaotic" version of the *whole* quantum mechanics, where *every* real, unreduced interaction or process results in explicit creation of dynamically redundant, probabilistic and intrinsically unstable (chaotically changing) structures [19,23-25]. Thus, *every* excited atomic state and excitation process appears to be dynamically chaotic (with a variable degree of randomness) and therefore fundamentally unstable and *only probabilistically* predictable in its detailed behaviour (in accord with the empirically based canonical postulates), whereas externally stable ground states result in reality from chaotic, permanently changing, but unresolved configurations of underlying interaction processes.

These fundamentally substantiated results of *intrinsically chaotic* quantum mechanics are directly applicable to the currently extremely popular idea of "quantum information processing" developed within usual, dynamically single-valued *projection* of real quantum system dynamics. It immediately follows from our unrestricted analysis [19] that physical realisation of *any unitary* quantum computation is *fundamentally impossible*, irrespective of details of its mathematical scheme or equally unitary ideas about its "stabilisation" ("chaos control") with respect to destructive "environmental influences". Indeed, any *single*, most elementary action of *real* computation by a most "isolated", or "controlled", system involves unreduced interaction leading to a change of system state (now specified as realisation or complexity level change) used in further computation actions. As follows from the nonperturbative interaction analysis, any nontrivial, significative interaction results in dynamically split products and each of them can be selected by the system, in an *irreducibly random* and *irreversible* (entropy-increasing) fashion. It is clear that such real quantum computer will quickly show appreciable deviations from its expected unitary computation scheme. Every attempt of "chaos control" is reduced to another interaction and therefore can only increase (but never eliminate) chaoticity and entropy growth, contrary to the idea about the possibility of "coherent" chaos control (or any real interaction dynamics) in unitary quantum computation theory.

It is true that the dynamic regime of "chaos control" type can provide high operation stability of ordinary, "classical" computers, but in that case one has a large "control space" due to the big difference between a characteristic "quantum" of complexity-action, $\mathcal{A}_0$, determining "classical bit" realisation and physically indivisible quanta determined by $h$, so that "controlling" influences can be much smaller than $\mathcal{A}_0 \gg h$, while still remaining much larger than $h$, which can give, in principle, a sufficient stability. In the world of *essentially quantum* phenomena, *everything* that happens is of the order of *minimum* complexity-action quantum $h$ or close to it, and "something" *dynamically unpredictable* will *necessarily* happen, as our analysis shows, within *every* real interaction, or elementary "action of computation", which devaluates any reasonable "control" [19]. Proposed applications of conventional quantum chaos approach to the analysis and control of quantum computer dynamics [60-62] only demonstrate once again the evident basic inconsistency of the concept of "chaos without chaos" that cannot be hidden behind formal reference to "quantum ergodicity" and other purely symbolical "tricks" of the unitary description, looking especially irrelevant after clear and extensive presentation of the causally complete, *dynamically multivalued* description of the *true* quantum chaos in real physical systems [19,23-25].

Note that one has an irreducible source of serious doubts in unitary quantum computation already within scholar quantum mechanics: all its "mysterious", formally postulated features could yet be practically ignored for the situation of single measurement over a simple quantum system, but if the detailed, multi-step dynamics of a sophisticated *and* essentially quantum system determines the desired results, as it is the case for quantum computers, then one should not be surprised to find some "unexpected" consequences of "mysteriously" hidden (from the *conventional* scheme), but actually *real* parts of the physically real system dynamics. Positive resolution of fundamental difficulties in realisation of unitary quantum computers is unambiguously suggested by the above "chaotic quantum mechanics": instead of basically inefficient "control of chaos" in the irreducibly chaotic device, one should use the "naughty chaos" for a good purpose and realise the unreduced, *dynamically multivalued* information processing within real, complex-dynamical quantum systems. However, any reasonable consideration of this idea is qualitatively, conceptually different from any unitary imitation by both dynamical mechanism and the very sense of "computation" and needs therefore the full application of causally complete, reality-based description of quantum dynamics clearly specified and substantiated within the quantum field mechanics [19,24,25]. This conclusion refers to *any* application of "detailed" quantum dynamics to a sufficiently complicated system, including the *majority* of currently developed "advanced" applications on essentially quantum scales ("nano(bio)technology", "molecular devices", etc.) remaining limited to the same basically deficient unitarity as usual quantum computers, which shows that the proposed causal completion of the standard Schrödinger formalism exceeds by far a purely academic interest.

### 3.2. Universal hierarchy of Schrödinger equations, wavefunction of the universe, and fundamental dynamical fractal

According to the previous section results, the extension of causally complete, dynamically multivalued version of Schrödinger dynamics includes not only the objectively understood content of "quantum strangeness" at the lowest complexity levels, but equally big generalisation of interaction dynamics at higher, "classical" levels, where causal wavefunction takes the form of complex-dynamic extension of classical "distribution functions". We have seen how dynamically multivalued, causally random interaction processes between "essentially quantum" systems like elementary particles lead to purely dynamic emergence of simplest classical, permanently localised objects, in the form of elementary bound systems like atoms. Complexity development does not stop there but further continues in the same way, based on universal dynamic redundance of interaction products and causal wavefunction role of their unification in indivisible complex dynamics. Thus, the obtained classical, intrinsically localised complex systems interact among them, and the emerging (classical) chaos of randomly changing realisations of a compound system endows its configuration, or (generalised) "trajectory", with an intrinsic "fuzziness" that can be relatively small or big depending on the interaction parameters. The generalised causal wavefunction, or distribution function, emerges here as the same physically real entity as the quantum-level wavefunction made up by permanent system transitions between its redundant realisations and having the same, dynamically probabilistic interpretation, according to causally extended "Born's probability rule".

A qualitative difference between the distribution functions of different levels can arise due to basically "wave-like" (distributed) or "corpuscular" (localised) nature of the main entities of respective complexity levels. Wavefunctions of "wave-like" levels tend to "real waves" producing interference patterns and give the probability density for observed realisation emergence in the form of their squared modulus. Wavefunctions of "corpuscular" levels describe the same distribution of realisation probability by their direct magnitude and cannot account for undular effects that do not appear at such corpuscular levels. However, this is not a major fact of fundamental importance for the whole picture. It can be consistently taken into account and does not change the main formalism of dynamic redundance by the extended EP method (section 2). Each of the qualitatively different types of behaviour and wavefunction can repeatedly occur at both "microscopic" and "macroscopic" levels of world complexity, and one may also have a level of complexity with an "intermediate" type of behaviour.

Universality of unreduced interaction analysis starting from eq. (1) shows that it can be directly applied to interactions at any level of complexity giving the same phenomenon of dynamic discreteness (quantization) of system realisations and "quantum jumps" between them, expressed by the same kind of causal quantization relations, eqs. (9)-(11), related to generalised causal wavefunction. The peculiarity of the very first, "quantum" level of complexity is that the characteristic quantum of complexity-action determining the "magnitude" of essentially nonlinear quantum beat cycle is strictly permanent and therefore fixed as a "universal" constant, Planck's constant $h$, which is due to the "indivisible" nature of this *lowest* complexity level [1,19]. The characteristic action quantum $\mathcal{A}_0$ at a higher complexity level has the same complex-dynamic interpretation, but may have one or several "effective" values that can be somewhat "smeared". Such more involved quantization structure at higher complexity levels is due also

A.P. Kirilyuk

to a more complicated form of the corresponding Hamiltonian, $H$, which, instead of simple case of eq. (12), is represented, in general, by arbitrary function of $x$ and $p$, $H = H(x,p)$, so that the "corpuscular" version of causally random system wandering over its redundant realisations given by eq. (12) for the elementary quantum system is extended to the *generalised Hamilton-Jacobi equation* [19]:

$$H\left(x, \frac{\Delta \mathcal{A}}{\Delta x}\bigg|_{t=\text{const}}\right) = E \;, \tag{22}$$

where we have substituted the universal expression of momentum as the spatial rate of complexity-action development, eq. (6). In the most general case of Hamiltonian explicitly depending on time, $H = H(x,p,t)$, the same equation, in combination with the universal energy definition as the temporal rate of action decrease, eq. (6), takes the form

$$\frac{\Delta \mathcal{A}}{\Delta x}\bigg|_{x=\text{const}} + H\left(x, \frac{\Delta \mathcal{A}}{\Delta x}\bigg|_{t=\text{const}}, t\right) = 0 \;. \tag{23}$$

The difference of this extended version of the Hamilton-Jacobi equation from the canonical one is not only in *dynamically discrete* derivatives replacing their usual, continuous limits, but especially in the revealed quantitatively new, *intrinsically chaotic* (multivalued) evolution of arbitrary system with interaction [19]. Permanent chaotic wandering of any system between its realisations, implied behind the generalised formalism, provides a causally complete extension of usual, abstract "least action principle" (and other "variational principles"), the related "Lagrangian" (function and formalism) and leads to the unified *Hamilton-Lagrange formalism*, including the "integral" version of eqs. (22), (23) [19]. It is that causally complete extension of usual, dynamically single-valued interpretation that permits one to derive the universal Schrödinger equation for the generalised wavefunction from eqs. (22)-(23). It is achieved with the help of mentioned generalisation of the causal quantization rule, eqs. (9):

$$\Delta \mathcal{A} = -\mathcal{A}_0 \frac{\Delta \Psi}{\Psi} \;, \tag{24}$$

where $\Psi$ is the generalised causal wavefunction (distribution function) probabilistically linking "regular" realisations (section 2), while the characteristic action quantum $\mathcal{A}_0$ may be additionally multiplied by $i$ for "wave-like" levels of complexity (cf. eq. 9(c)). Now the universal Schrödinger equation is obtained by substitution of expression of eq. (24) for $\Delta \mathcal{A}$ in eqs. (22) and (23), where one should actually take into account possible variations of $\mathcal{A}_0$ for different terms in the Hamiltonian:

$$\hat{H}\left(x, \frac{\partial}{\partial x}\right)\Psi(x) = E \;, \tag{25}$$

$$\mathcal{A}_0 \frac{\partial \Psi}{\partial t} + \hat{H}\left(x, \frac{\partial}{\partial x}, t\right)\Psi(x,t) = 0 \;, \tag{26}$$

where the "operator" version of Hamiltonian function, $\hat{H}$, takes into account causal quantization rules of eq. (24) and their proper raising to power (increasing the derivative degree) [19-22], $x$ stands for the *causally emerging* system configuration (section 2.2), and derivatives should be understood, in general, as their dynamically discrete versions.

Particular cases of universal Schrödinger equation are obtained from its general form, eq. (26), for the corresponding particular forms of Hamiltonian functional dependence, $H(x,p,t)$. Expanding this function in a power series of $p$ (where coefficients can depend upon and be additionally expanded in power series of $\mathcal{A}$ or $\Psi$), one obtains the following universal Schrödinger equation

$$\frac{\partial \Psi}{\partial t} + \sum_{n>0, m} h_{mn}(x,t)[\Psi(x,t)]^m \frac{\partial^n \Psi}{\partial x^n} + \sum_m h_{m0}(x,t)[\Psi(x,t)]^{m+1} = 0 \;, \tag{27}$$

where $h_{mn}(x,t)$ are dynamically determined coefficients of Hamiltonian expansion (including generalised "interaction potentials" and "masses") and summation over integers *n* and *m* can include any intervals of their values between 0 and ∞, but is usually limited to a small number of lower values. It can easily be seen that this equation covers practically all known equations for "distribution functions" and "order parameters" accounting for structure formation processes (see e. g. [63]), even though not all of those "model" equations will necessarily correspond to a realistic, causally extended version of eqs. (25)-(27) and not all real cases need to be described by a Hamiltonian power-series *expansion* of eq. (27). In this sense, the universal Schrödinger equation, eq. (26), together with its dual, "corpuscular" companion, the generalised Hamilton-Jacobi equation, eq. (23), can be considered as the *unified Hamilton-Schrödinger formalism* incorporating the whole diversity of all known (and unknown) particular equations.

It is important to stress that the extension obtained is not a formal mathematical generalisation of particular equations. As we have shown above, the Schrödinger equation describing system complexity development at its certain level is causally derived from the unreduced analysis of interaction between entities of lower level(s), which always includes the dynamic redundance and entanglement phenomena totally absent in canonical science that simply postulates, in each particular case, a "suitable" form of dynamic equations and tries to mechanically fit their severely limited, dynamically single-valued solutions to observation results. That blind trial-and-error search and purely technical tricks of usual empiricism lead to abuse of "mysteries" and abstraction already at the lowest complexity level of the unified world dynamics and to practical absence of any objective description for higher levels of complexity, where the number of necessary "postulates" grows dramatically, in proportion to realisation number determining system complexity. Rigorous derivation and complex-dynamic interpretation of universal Schrödinger formalism provide, in particular, the inherent *creativity* and *dynamic adaptability* for the emerging structures that make them realistically "alive"/self-developing, while being absent in unitary projections of scholar science, those properties should be artificially added in an inevitably incomplete form.

The universal Schrödinger equation is provided with the equally universal method of its causally complete analysis within the unreduced EP formalism (section 2) giving its complete *general solution* in the form of dynamically probabilistic realisation sum and probability distribution (eqs. (2)-(3), see refs. [2,19-22] for detailed expressions). Due to dynamically redundant and internally entangled nature of the general solution of universal Schrödinger equation, one can often limit the expansion of eq. (27) to several lowest powers, since when the "ordinary", non-dynamic nonlinearity increases with power, the system actually passes to a higher sublevel of complexity through emergence of new structural elements whose interaction is described again by a lower-power, often formally linear Schrödinger equation. Natural creativity (dynamic instability) of unreduced interaction processes tends thus to "automatically" *simplify* the initial, formal interaction configuration, providing a complex-dynamical version of "Occam's principle of parsimony", while the true complexity of observed richness of forms and behaviour results from dynamic redundance and fractal entanglement of products of that formally "simple" interaction, as they naturally emerge in its *unreduced* analysis. In other words, it is the *essential, dynamic nonlinearity* arising from the natural feedback loop formation of unreduced interaction process (section 2.1) that is at the origin of real structure formation, rather than its unitary imitation by the false "nonlinearity" of usual theory reduced to fixed, mechanistic "curvature" of given external forms. The described creative type of real system evolution explains also the real meaning and relative efficiency of unitary science postulates limiting the maximum power of various purely abstract terms in basic structures of the canonical theory (Lagrangian, etc.). Now we see, however, that the detailed Hamiltonian structure should be determined for each particular case, starting from complex-dynamic interaction development itself, which excludes any universality of *details* (and actually leads to the observed *diversity* of interaction products in various particular cases). It means that the effective Hamiltonian structure of universal Schrödinger formalism, eqs. (25)-(27), "automatically" changes itself, in accord with progressively emerging structures of each next (sub)level of complexity, which is another manifestation of intrinsic *creativity* of the extended description, totally absent in its canonical, dynamically single-valued projection.

Universality of the extended Schrödinger formalism means also that many particular features of the causally complete, multivalued dynamics, usually attributed to "inexplicable", specifically "quantum" properties in the case of micro-system behaviour, will reproduce themselves at higher, "classical" and "macroscopic" levels of complexity. Such is, for example, the basic property of dynamic discreteness (quantization) of any real interaction dynamics that leads to its irreducibly nonuniform, "step-like" character. If in the special case of large number of similar system components taken at the end of their complexity development process (i. e. in the state of generalised system *equilibrium* [19]) those dynamic inhomogeneities can have a relatively fine-grained structure of small-scale fluctuations, in the general case of intense complexity development, step-like emergence/extinction events, or "huge fluctuations", will be a "standard", inevitable manifestation of unreduced interaction dynamics, remaining "mysterious" for unitary science or inconsistently attributed to equally big *external* influences. A related "quantum" property of any complex interaction dynamics is expressed by "dynamic uncertainty relations", which are none other than reflection of the same dynamic discreteness property (see eqs. (5), (6)) at any level of complexity [19]. The universal "coordinate-momentum uncertainty relation" means, in particular, that the extended distribution function for *any* system, analogous to the quantum-mechanical wavefunction, cannot simultaneously depend on *exact* values of both coordinate and momentum, as opposed to the accepted convention for classical systems. The difference will be essential in the case of non-equilibrium systems representing most interesting cases of *actually changing* reality, which implies the necessity of fundamental extension of basically perturbative versions of usual "kinetic" theories.

Note an essential difference between our causally complete analysis and empirically based imitations of *dynamic uncertainty* in scholar, dynamically single-valued "science of complexity" [39]: in such unitary imitations of complexity the *postulated* "fuzziness" of system behaviour is *not* obtained from a consistent analysis of system dynamics, while the "distribution function" of a "chaotic" system is supposed to be a (dynamically single-valued) solution of a formal generalisation of a regular distribution-function formalism, such as Liouville equation, providing a clear example of mechanistically reduced interaction analysis. That imitation of unitary "science of complexity" is close to that of conventional "density matrix" approach in quantum theory described above (section 3.1).

The obtained complex-dynamic hierarchy of particular realisations of the universal Schrödinger formalism and causally extended wavefunction for the respective, really observed levels of complex dynamics provides also the causally complete solution to one of the most intriguing and fundamental "mysteries" of canonical theory, that of the *"wavefunction of the universe"*. Not only unitary theory cannot develop any consistent understanding of the wavefunction and related quantum behaviour at the lowest levels of world dynamics, but it can neither explain how the wavefunction of primordial, quantum state of the universe is transformed into the observed diversity of apparently "incoherent" and "non-quantum" forms, while the quantum-mechanical wavefunction preserves its dominance at the most fundamental levels of being. The problem can be considered also as that of *classicality emergence* on the scale of the *whole universe*, which provides additional difficulties for canonical theory, since on that scale it cannot, even formally, apply its usual trickery of particular quantum system "decoherence" because of *external* "influences" magically transforming it into a classical one (unless one accepts a yet more doubtful assumption about an omnipresent "decohering" influence coming to our world from some "outside" reality and "choosing", in a very peculiar fashion, the systems to be made "classical").

We have shown above how the problems of probabilistic and configurationally changing wavefunction can be causally resolved within its unreduced, complex-dynamic interpretation as *transitional* state-realisation of a dynamically multivalued system, for both low, "quantum" and higher, "classical" levels of complexity. We have also provided the causally complete, purely dynamic, internal origin of classicality describing it as a "generalised phase transition" [19] between complexity levels starting generically from elementary bound system formation. It is even more important, however, that all those systems from various naturally emerging levels of complexity remain *permanently and physically connected* within the *unified* dynamics of unceasing development of fundamental interaction in the initially homogeneous system of two coupled protofields (section 2.1). This fact, confirmed by the *direct derivation* of multivalued interaction dynamics at each complexity level from the products of lower-level complexity development within the *same, universally valid* formalism, means that we deal indeed with the *unified*, hierarchically structured, and totally *realistic wavefunction of the universe* physically represented by various coupled protofield perturbations (and perceived rather at the e/m side of the system) that form the observed world structures, in their full diversity of forms and behaviour. "Decoherence" of observed macroscopic structures naturally results from mutual *independence* and *spatially chaotic* emergence of locally "complete", incompatible system realisations (it is rather the *coherence* of certain "quantum" cases that is provided with a *special* explanation of increased synchronisation, or "complex-dynamic self-organisation", in the corresponding regimes of multivalued dynamics [19]). That typical absence of "quantum coherence" between world structures and related independence of their local dynamics does not enter in contradiction with their *intrinsic unification* in the holistic complex dynamics of the world. Such possibility is due to the dynamic multivaluedness phenomenon itself, naturally emerging in the *physically unified* world construction of the two coupled protofields, so that neither the observed world diversity, nor its indispensable unity needs to be sacrificed. Observed manifestations of real unity of world dynamics include universality of two "macroscopic" interactions, e/m and gravitational, and main physical constants, remaining unexplained in usual science, including the most "elegant" imitations of modern field theory, and provided now with the *causally complete* explanation within the dynamically multivalued description of protofield interaction development (section 2) [1,2,19-22].

Having obtained dynamic unification and corresponding mathematical universality of the causally extended wavefunction, we can now apply that universal description to an arbitrary level of complexity, scale, system, or stage of its complex-dynamical development. Thus, the "quantum" stage of the world structure emergence starts from formation of elementary particles provided with the causally extended (universal and individual) wavefunctions, as it is described by the existence equation analysis and resulting extended Schrödinger formalism (section 2.1) presented here only in outline and supposing further development by the same method. It leads, in particular, to the causally extended version of "quantum" and "classical" *cosmology* containing serious, qualitative modifications with respect to existing unitary versions and their consequences. On the other hand, the same causal wavefunction concept and extended Schrödinger formalism can be applied to *any* system of interacting "classical" objects, providing the general, causally complete solution to the many-body problem in the same form of causally probabilistic realisation sum, eqs. (2)-(3), which extends qualitatively potentialities of classical and quantum description of many-body systems fundamentally limited, in its usual version, to the same dynamic single-valuedness of any its perturbative solution [19].

The emerging complex-dynamical hierarchy of systems, their realisations and wavefunctions described by the corresponding hierarchy of particular cases of universal Schrödinger formalism, eqs. (25)-(27), can also be presented as dynamically extended, *intrinsically probabilistic* and *self-developing* version of *fractal* representing now the whole, real world structure [19]. That fundamental dynamical fractal of the world represents the *totality* of its contents, with all its structures and their dynamical regimes, and not only a mechanically simplified, regular and fixed (dynamically single-valued) imitation of certain types of structure provided by usual, mathematical fractals [64-67]. Moreover, even each part of the unified world fractal, corresponding to a particular system or level of complexity, is different from canonical fractality due to permanently and chaotically changing, dynamically probabilistic structure of extended fractality determined by the dynamically multivalued origin of any its element. This property of real structures can only be obtained within the unreduced, universally nonperturbative analysis of underlying interaction processes, and such interaction analysis is absent in usual fractality concept.

The extended dynamical fractal of the world unifies hierarchically structured causal wavefunctions and realisation sets for various systems and levels of complexity, so that the "wavefunction" as such forms the "infinitely" fine "fuzziness" of the corresponding level of the world fractal, while "regular" realisations connected by that fractal net (and usually gathered in dense groups) appear as finite "branches" of the fractal, or "objects", at the same level of complexity. The unreduced dynamic complexity of the first, relatively "fine" part of the real world fractal can be interpreted as complex-dynamical version of "potential energy" of interaction between "distinct", rough objects (grouped realisations) called also *dynamic information* [19], while moving objects themselves contribute to the "generalised kinetic energy", or *dynamic entropy*. Both "information" and "entropy" represent integral parts of the same fractal structure of complexity, corresponding to the two its possible forms, less stable "potential" form (information) and more stable "structural" form (entropy). Whereas the sum of information and entropy, constituting the total dynamic complexity of a system (e. g. the world) is always conserved, each of its two parts changes unceasingly, so that dynamic information always decreases, while dynamical entropy increases, which is the rigorous, unified expression of interaction (complexity) development process described above. This *universal law of complexity conservation and development* unifying the extended versions of the first and second laws of thermodynamics (conservation and degradation of energy respectively), as well as all other fundamental laws and "principles" of canonical science, reflects permanent, causally probabilistic change (self-development) of the fundamental dynamical fractal of the world, and the above universal Schrödinger formalism, eqs. (25)-(27), and its "corpuscular" complement of extended Hamilton-Lagrange formalism, eqs. (22)-(23), can be derived as the unified basic expression of that universal evolution law [19].

The fractally structured, hierarchical wavefunction of the universe giving realisation probability distribution at each level of complexity is the natural unifying entity of complex world dynamics reflecting the physically unified structure of two interacting protofields. Such is the unreduced result of the truly causal, ultimately complete extension of the wavefunction concept introduced by Erwin Schrödinger 75 years ago and crudely, fundamentally reduced by conventional, "mathematical" physics to over-simplified, purely abstract "vectors" in purely abstract spaces that cannot consistently describe even the lowest level of world dynamics. The causally extended wavefunction of quantum field mechanics reproducing *exactly* properties of real entities is obtained as a result of *elimination of all artificial limitations* of canonical, dynamically single-valued (unitary) description and *freedom* for the underlying *interaction process* to develop in its natural, unlimited, universally nonperturbative way. This approach suggests the explicit, progressive *derivation* of naturally emerging entities of *all* levels of complexity and their causal wavefunctions, in *full correspondence* with their *real creation* by *dynamically continuous*, though quite non-uniform (non-unitary), *entanglement* of interacting entities from lower levels, which permits us to avoid arbitrary, non-causal "extensions" of certain esoteric "generalisations" of quantum mechanics that tend to postulate direct, inconsistently reductive links between the lowest, quantum-mechanical and highest levels of complexity accounting for brain activity and the property of consciousness (see e. g. [68-70]). Brain operation can indeed be causally described within the complex-dynamic wavefunction formalism, including the property of (genuine) consciousness that emerges starting from certain, high enough level of universal dynamic complexity [19], but that causally complete description of *highest*-level complexity properties deals with the dynamically multivalued interaction of *physically real* entities from the *corresponding* complexity levels (electro-chemical interactions between brain elements), so that "quantum-like" features of brain dynamics emerge rather due to the obtained *universality* of unreduced, multivalued interaction dynamics and its causal wavefunction. The latter can also be called, in the case of brain dynamics, the *brainfunction* and describes the dynamically *quantized*, causally *probabilistic* wave field configurations of electro-chemical interactions of neurones and their groups, *unifying* the hierarchy of dynamically redundant, "localised" *realisations* (that correspond to "thoughts", "impressions", etc.) within the *self-developing*, intrinsically *creative* brain dynamics. That highest perceived level of the universal wavefunction, not restricted from above, demonstrates the true, already attained potentialities and unlimited perspectives for further development of the causally extended Schrödinger picture and formalism presented above.

## 4. Conclusion: Dynamically multivalued wave mechanics and reunification of causally extended new physics by the universal concept of complexity

A strange similarity in the way of emergence of three major discoveries of basic quantum principles at the beginning of 20th century, introduction of quanta by Max Planck [1], of matter wave by Louis de Broglie [2], and of the wavefunction and its equation by Erwin Schrödinger (section 1), can be seen now as manifestation of a profound unifying basis of those revelations and underlying attitudes of their authors, differing them from other directions in the "new physics" development of the same period. It is certainly the unreduced logical causality, integrity and physical realism that unify the three founders of quantum mechanics and separate them from the glaring inconsistency of the opposite, "math-physical" way of development, artificially and subjectively imposed to physics (and science in general) during its later evolution in the 20th century. Although Planck, de Broglie, and Schrödinger clearly opposed themselves to the growing domination of inconsistent mathematical abstraction, certain superficial, quantitative correlations between the canonical formalism results and experimental data, combined with trickily exploited real successes of technological, *purely empirical* progress, have led to general acceptance of "mathematical" way of science development, despite its *evident* flaws and clear objections of many great creators of "new physics". As a result, the "corner of the great mystery" unveiled by them at the beginning of 20th century remains without change until now, despite so many efforts, glorification and huge "computer power" of unitary science applied during the century of "scientific revolutions".

Basic deficiency of canonical science and destructive consequences of its artificially maintained, mechanistic paradigm become the more and more evident, in both their fundamental and practical aspects and for various fields of knowledge [1,19,22,71]. It is clear that the dangerously degrading situation cannot be changed without *explicit* introduction of a *qualitatively new*, well-specified *paradigm of knowledge* of a *superior* consistency that could demonstrate its efficiency by *direct solution* of stagnating problems and should necessarily involve the *content* of knowledge and not only its form or way of public presentation (as it is often proposed today as an "easy" way to improve the situation in science without changing its essential content). In this and other works [1,2,19-22] we have presented such new form of scientific knowledge, the "universal science of complexity" based on the consistently derived "dynamic redundance paradigm", and demonstrated the resulting explicit solutions of the main problems of fundamental physics and other fields of science. In that way, the tendency for irreducible causality, realism and universality of knowledge of the three great creators of quantum mechanics finds its natural continuation and most complete realisation. We show that the qualitatively new feature, just missing (though often intuitively guessed) in the original "new physics" and now providing a reasonably big price for its causal completion is the unreduced *dynamic complexity* of *any* real *interaction* process determined by the naturally emerging phenomena of *dynamic redundance and entanglement*, which are properly obtained in the *universally nonperturbative* description of the new theory. It becomes clear that the observed unified reality *is* the unreduced dynamic complexity thus interpreted (universal interaction development by dynamic redundance and entanglement of hierarchically emerging entities), which provides the *rigorously specified* definitions of "realism" and "wholeness".

The causally extended wavefunction presented in this paper is the physically and conceptually unifying entity of the new, totally realistic description of "quantum" (and "classical") dynamics, since it actually links together and "manages" the chaotic mixture and alternation of complementary components and aspects of unreduced, multivalued dynamics. Extended Planck's "quantization" of fundamental protofield interaction into "corpuscular", dynamically squeezed realisations and de Broglie's "matter wave" formation between those localised realisations are naturally unified and governed by that intrinsically chaotic (and therefore always *partially* ordered), dynamically probabilistic intermediate state of system jump-like transitions between realisations. And although the formal absence of the source of complexity (dynamic redundance) in the dominating unitary paradigm of conventional science did not permit to the true founders of quantum mechanics to find a causally complete explanation for the peculiar properties of quantum behaviour, each of them preferred to continue, often desperately, the search for such causal explanation, rather than complaisantly yield to the dominating mystification of "standard" theory. Recall, for example, the definite rejection by Schrödinger of "quantum jumps" *in the absence of their causal substantiation* [14] that can only now be *uniquely* obtained within the dynamic redundance paradigm, in contrast to multiple basically deficient *imitations* of "real collapse" of *abstract* "state vector" in the dynamically single-valued approach inevitably involving *evidently inconsistent* modifications of Schrödinger formalism. In our theory we *rigorously derive* instead the *exact* form of usual Schrödinger equation *as a result* of *causally specified* "quantum jumps" (of virtual soliton).

Another line of development of standard quantum mechanics leading to ever growing abstraction and pseudo-philosophical, verbal "games with reality" starts from the "competing" scheme of quantum behaviour, its "matrix" version of Heisenberg, Born, and Jordan (see e.g. ref. [72] for detailed story and references). Although it is usually considered that Schrödinger has demonstrated, in one of his pioneering papers of 1926 [10], the equivalence of the two descriptions, it would be more appropriate to say that he showed only that the previously published "matrix" version can be obtained from his *Wellenmechanik*. As to the reverse relation, one can immediately notice that it is the main, "reality-oriented" entity, the wavefunction itself, that is basically missing in the demonstratively "mathematical" scheme of "matrix mechanics". We can see now, within our causally complete version of quantum behaviour, that the "matrix" approach emphasises, though in a mechanically reduced, abstract form, a physically real property of unreduced dynamic complexity, dynamic discreteness (quantization) of any real interaction process, but contrary to the wavefunction, the "matrix" description, as if generalising the "purely mathematical" discreteness of integer numbers, inevitably tends to definite "liberation from reality", in favour of purely abstract, *incredibly* simplified "spaces" of unitary and now unfortunately dominating "mathematical physics". Contrary to Schrödinger equation, "matrix equations" do not directly describe evolution of any intrinsic, physically relevant and "global" quantity of the whole system representing the essence of its dynamics. We have shown, within our complex-dynamic description, how the wavefunction can indeed be provided with the direct realistic interpretation of a "wave field" (due to the *dynamic chaoticity* of the latter). As for "matrices" and "operators", a direct realistic interpretation remains impossible for them, even though they represent understandable, but very rough, simplification of real complex-dynamical processes. If a "matrix" can be generated from the wavefunction (in the form of "matrix elements" of interaction potential expressing a sort of "interaction-driven transition" between two wavefunction states/realisations), the reverse is hardly possible; causal wavefunction could be approximately described as a complicated, chaotically changing "knit" product "woven" from various possible matrix elements, but it is not really a consistent idea.

In any case, the tendency of purely abstract science has made a big "progress" since 1926, so that its modern version of "string theories", in order to find a strong financial support and be considered the most promising candidate for the "final theory of everything", should not even bother any more about any relation to reality, replacing it with a self-generated, subjective estimate of internal "elegance" always confirmed by the respective "international community". That they leave without answers all the canonical "quantum puzzles" (as well as "relativistic paradoxes" and other unsolved problems of "new physics"), that the announced "unification" is a purely abstract game of dead, trickily "fitted" symbols, that they create a growing number of new inconsistencies within them and with respect to observations — all these inherent properties of "advanced-study" theories of "mathematical" physics do not really matter for self-"chosen" sages living in their "abstract spaces" (but consuming real-space resources) and *always* knowing better what is elegant or ugly *for them*.

In the meanwhile, the unique, unreduced reality wins everywhere in real world spaces. Not only it wins when the "sages" descend from their towers of pure mathematical thought to buy real products in a real supermarket that cannot be described in principle, in any part by their over-simplified abstractions. It wins also by already large and still growing discrepancies between the predictions of somewhat "too symmetric" theories and observations over our real, so "asymmetric" world (which some "ordinary" people consider to be beautiful just *due* to its *higher*, "irregular" kind of symmetry). And finally, the unreduced reality wins by a "strangely" dropping interest of young people to abstract science, despite multiply increased accessibility, diversity and intensity of information (not to say "agitation") about incredibly "promising" and excitingly "dramatic" development of canonical theories. Maybe it is so because the unreduced human mind, being a complex-dynamical system of a superior complexity level, still instinctively resists the imposed canonical abstraction that has strictly zero dynamic complexity, as it follows from the proposed universal description of the latter.

With the appearance and detailed presentation of unreduced, complex-dynamical description of quantum and higher-level processes [19] not only the glaring flaws of unitary theory become yet more evident (for now we specify the *exact, universal reason* for its omnipresent divergence from reality), but one obtains also an *explicit, well-specified* realisation of qualitatively new, *intrinsically complete* kind of theory that indeed *resolves* explicitly all the main problems of canonical description, and this is just due to the properties provided by extension from a dynamically single-valued projection to the unreduced, dynamically multivalued version of reality. That crucial transition is obtained as a result of definite *elimination of artificial limitation* of natural *interaction development*, which *should* give the adequate picture of reality and corresponds to its *truly* exact and rigorous description, as opposed to *false* "exactness" and "rigour" of canonical, dynamically single-valued theory. We have also demonstrated (see especially refs. [1,19]) how the causally complete, complex-dynamical description of elementary particles and their interactions helps to resolve many *practically important problems* in experimental research strategy and applications led to a fundamental impasse by the imposed unitary approach results. As an example, we can mention here applications of dynamically multivalued theory of quantum chaos and quantum measurement, including quantum computers and other quantum devices (section 3.1).

That crucial, causally complete extension of fundamental understanding of reality and its practical development seems to be much more pertinent for celebration of new physics jubilees and science "millennium boundaries", than official "events" and "discussions" filled up with dull repetitions of the same, always *unsolved* problems only confirming the impotence and futility of unitary science (see e. g. [73-77]). What they really "celebrate" is the *death* of *that* kind of science. While the continuing purely subjective, unreasonable and unfair neglect by stagnating canonical science of emerging causally complete solutions of fundamental and practical problems can only be attributed to "scientific revolution effects" emphasised by T. Kuhn [78] — that is to occupation and preservation of highest "official" positions in science at any price and especially against the "pressure" of unreduced truth — it becomes clear that the century of intellectual and moral degradation is now definitely behind us, and further domination of trickery and manipulation in fundamental knowledge can only create increasingly serious problems for their authors and danger for civilisation development. By contrast, a timely, and now quite realistic, transition to the evidently advantageous, complex-dynamic description of complex reality, avoiding any artificial imitation and simplification, will create a new, powerful source of progressive development of fundamental science replacing its current state of decay that marks the definite end of the whole unitary kind of knowledge.